\documentclass[12pt,preprint]{aastex}
\usepackage{graphicx}

\shorttitle{Chemistry in Disks}
\shortauthors{Vasyunin et al.}

\begin{document}

\title{Chemistry in Protoplanetary Disks: A Sensitivity Analysis}

\author{A.I. Vasyunin}
\affil{Max Planck Institute for Astronomy, K\"onigstuhl 17,
D-69117 Heidelberg, Germany}
\affil{Ural State University, ul. Lenina 51, Yekatirinburg, 620083
Russia}
\email{vasyunin@mpia.de}

\author{D. Semenov, Th. Henning}
\affil{Max Planck Institute for Astronomy, K\"onigstuhl 17,
D-69117 Heidelberg, Germany}
\email{semenov,henning@mpia.de}

\author{V. Wakelam}
\affil{Universit\'e Bordeaux 1, CNRS, Laboratoire Astrophysique de Bordeaux , BP89 33270 Floirac, France}
\email{Valentine.Wakelam@obs.u-bordeaux1.fr}

\author{Eric Herbst}
\affil{Ohio State University, Department of Physics, 174 West 18th
Avenue, Columbus, OH 43210-1106 USA}
\email{herbst@mps.ohio-state.edu}

\and

\author{A.M. Sobolev}
\affil{Ural State University, ul. Lenina 51, Yekatirinburg, 620083
Russia}
\email{andrej.sobolev@usu.ru}

\begin{abstract}
We study how uncertainties in the rate coefficients of chemical
reactions in the RATE\,06 database affect
abundances and column densities of key molecules in protoplanetary
disks. We randomly varied the gas-phase reaction rates within their
uncertainty limits and calculated the time-dependent
abundances and column densities using a gas-grain chemical model and a flaring steady-state
disk model. We find that key species can be separated into two distinct groups according
to the sensitivity of their column densities to the rate uncertainties. The first group
includes CO, C$^+$, H$_3^+$, H$_2$O, NH$_3$, N$_2$H$^+$, and HCNH$^+$.
For these species, the column densities are not very
sensitive to the rate uncertainties but the abundances in specific regions are.
The second group includes CS, CO$_2$, HCO$^+$, H$_2$CO, C$_2$H, CN, HCN, HNC
and other, more complex species, for which high abundances and abundance uncertainties co-exist in the same disk
region, leading to larger scatters in the column densities.
However, even for complex and heavy molecules, the dispersion in their column densities is not more than a factor
of $\sim 4$. We perform a sensitivity analysis of the computed abundances to rate uncertainties and identify
those reactions with the most problematic rate coefficients. We conclude that the rate coefficients of
about a hundred of chemical reactions need to be determined more accurately
in order to greatly improve the reliability of modern astrochemical models.
This improvement should be an ultimate goal for future laboratory studies and theoretical
investigations.
\end{abstract}

\keywords{accretion disks -- astrochemistry --- methods: statistical ---
molecular processes --- stars: planetary systems: protoplanetary disks}

\section{Introduction}
More than 140 organic and inorganic molecular species consisting of up to 13 atoms
have been identified in space so far\footnote{\url{http://astrochemistry.net}},
including such complex species as dimethyl ether \citep[CH$_3$OCH$_3$;][]{Snyder_ea74}
and acetamide \citep[CH$_3$CONH$_2$;][]{Hollis_ea06}. The rich variety of the observed
molecules implies that many more relevant yet undiscovered species
must be involved in the processes of their formation and destruction. Still, only a tiny
fraction of these species have been firmly detected in several protoplanetary
disks, including CO (and its isotopes), CN, HCN, HNC, H$_2$CO, C$_2$H, CS,
HCO$^+$, H$^{13}$CO$^+$, DCO$^+$, and N$_2$H$^+$
\citep{Dutrey_ea97,Kastner_etal1997,Zadelhoff_etal2001,Qi_etal2003,Thi_etal2004}.
Multi-molecule, multi-transition observations of emission lines with (sub-)millimeter
interferometers and single-dish antennas provide a wealth of information about planet-forming disks resembling
the young Solar Nebula \citep{Simon_ea00,Aikawa_ea03,Qi_ea05,Pety_ea06,Dutrey_ea07}.

An ultimate goal of  disk studies is the reconstruction of the
evolutionary history and spatial abundance distribution of various
species, which requires sophisticated chemical models
\citep{Willacy_Langer00,Markwick_ea02,van_Zadelhoff_ea03,Ilgner_ea04,Semenov_ea05,Aikawa_ea06,TG07}.
The analysis and modeling of observational data acquired with
limited spatial and spectral resolution is not complete without
taking all possible uncertainties into account such as instrumental
errors, uncertainties in the distance to source, orientation, etc.
Whereas these uncertainties can in general be reduced either by
using better observational data provided by current interferometers
(PdBI, SMA) equipped with a new generation of receivers or in the
future with the advent of more powerful instruments like ALMA,
EVLA, and Herschel, there is an \emph{intrinsic} source of
ambiguity in all the chemical models -- our limited knowledge of the
reaction rate coefficients.

Modern astrochemical databases include up to about 4\,500 gas-phase reactions and $\sim 450$
species \citep{umist95,umist99,OSU03,rate06}, but many of these reactions have
poorly estimated rate coefficients with uncertainties of about a factor of 2 and larger.
For example, radiative association reactions lead to the formation of new complex species out of
smaller ones through the photon relaxation of an excited collisional molecular complex.
Under low-density space conditions where three-body processes are unlikely,
these reactions may have rate coefficients as high as $\sim 10^{-9}$~cm$^3$\,s$^{-1}$ and as
low as $\sim 10^{-17}$~cm$^3$\,s$^{-1}$, depending on the density of vibrational states of the complex.
The radiative association rates are usually difficult to measure in the laboratory as well as to predict
theoretically, especially for bigger species \citep[][]{Bates_51,Williams72,Herbst_80}.

In contrast, dissociative recombination of molecular ions almost always proceeds very rapidly, in particular
at low temperature, with rate coefficients that can be accurately
obtained \citep{Florescu_Mitchell06}. However, the products and their relative branching fractions branching
channels of many
dissociative recombination reactions are not known precisely, particularly at low temperatures
\citep[see though][]{Semaniak_ea01,Ishii_ea06,Ojekull_ea06}, which may spoil theoretical predictions
\citep[see, e.g.][]{Millar_ea88,Geppert_ea05a}.

Recently, \citet[Paper~I hereafter]{Vasyunin_ea04} investigated the influence of  uncertainties in rate coefficients
on  molecular abundances in dense and diffuse clouds, using the gas-phase UMIST\,95
database. They found that the abundance uncertainties of simple species are limited to within about an order of
magnitude and increase substantially with the number of atoms in the molecule, though the uncertainties still
do not exceed the observational errors for simple molecules. They  proposed a
sensitivity analysis to identify those reactions that introduce the largest errors
in the computed concentrations. \citet[hereafter Paper~II]{Wakelam_ea05}
studied uncertainties in the gas-phase chemistry of hot cores and demonstrated that at late times, $\ga 10^4$~yr,
modeled abundances of important molecules can suffer from large uncertainties due to poorly known rate coefficients.
\citet{Wakelam_ea06}  focused on  dense cloud chemistry and took uncertainties in the physical
parameters into account, as well as comparing the osu.2003 and UMIST\,99 chemistries. These authors came to
conclusions that grain-surface reactions should be considered in order to achieve better consistency with the
observational data, even though there is a large degree of ambiguity in simple cloud models based on static
physical and chemical structures. Finally, \citet{Wakelam_ea06b}  expanded this study by using their uncertainty analysis to show that  there is a bistability in the abundances of
many species that are hyper-sensitive to the adopted value of the ratio of the cosmic ray ionization rate of helium to that of hydrogen
\citep[see also][]{Pineau_des_Forets_ea92,Shalabiea_Greenberg95,Boger_Sternberg06}.
Furthermore,  \citet{Izzard_ea07} recently studied the effect of proton-capture reaction rate uncertainties
in the NeNa and MgAl chains on the abundances of the Ne, Na, Mg and Al isotopes produced in  intermediate-mass
AGB stars.

Despite previous work in rate uncertainties, to the best of our knowledge, there have been no attempts in the
literature to study how these  reaction
uncertainties, which are unlikely to be eliminated in the near future,
affect the results of astrochemical modeling in the wide range of
physical conditions typical of protoplanetary disks.  In this paper,
we extend the previous analyses by \citet[Paper~I]{Vasyunin_ea04} and
\citet[Paper~II]{Wakelam_ea05} to the conditions of a low-mass protoplanetary
disk surrounding a young T Tauri star. The aims of our study are
several. First, we analyze how large the abundance and column
density scatter for key species are due to uncertainties in rate coefficients.
Second, we investigate
how these uncertainties vary with disk location.
Third, we isolate the reactions whose rate
uncertainties contribute most to the abundance scatter and which are therefore worth studying in detail. Last,
we predict how the overall consistency of theoretical models will be improved
after these reaction rates are better constrained.

The organization of our paper is the following. The adopted disk
physical structure and time-dependent gas-grain chemical model are
presented in Sect.~\ref{phys_par}. The Monte Carlo method used to introduce the
uncertainties in the RATE\,06 reaction rates is described in Sect.~\ref{method}.
In Section~\ref{res} we report computed distributions of the mean molecular
abundances and column densities and their errors in the whole disk, and analyze
the influence of the rate coefficient uncertainties on these quantities. A correlation
method that allows determination of the role of various reactions in the abundance uncertainties
as a function of time and disk location is outlined, and results for several key species
are presented in Sect.~\ref{corr_method}. Particular attention is paid to the identification
of the most uncertain reactions for the chemical evolution of key species in the entire disk.
In Sect.~\ref{cons_impr} we predict to what extent the abundance scatter due to the rate
uncertainties will be lowered when accurate rate values of several dozens most problematic reactions
are available. The remainder of Section \ref{diss} is concerned with problems in different classes of reactions.
Final conclusions are drawn in Sect~\ref{concl}.

\section{Disk model and uncertainty approach}
\label{mod}

\subsection{Physical structure and chemical model of the disk}
\label{phys_par}
In our simulations, we adopted the 1+1D steady-state irradiated disk model with vertical
temperature gradient that represents
the low-mass Class~II protoplanetary disk surrounding the young T Tauri star DM Tau \citep{DAlessio_ea99}.
The disk has a radius of 800~AU, an accretion rate $\dot{M}=10^{-8}\,M_\sun$\,yr$^{-1}$, a viscosity
parameter $\alpha = 0.01$, and a mass $M\simeq0.07\,M_\sun$ \citep{Dutrey_ea97,Pietu_ea07}.
The thermal and density structure of the disk is shown in Fig.~\ref{disk}.
As a disk age we used a value of $\sim 5$~Myr, which has been derived by \citet[][]{Simon_ea00}
based on evolutionary track modeling of the central star. 
In chemical simulations an outer disk region beyond the distance of 50~AU from the central star
is considered. This is the only routinely accessible disk region with existing (sub-)millimeter interferometers.

We assumed that the disk is illuminated by UV radiation from the central star with
an intensity $G=410\,G_0$ at $r=100$~AU ($G(r) \propto r^{-0.5}$) and by interstellar UV radiation with
intensity $G_0$ in plane-parallel geometry \citep[][]{Draine_78,van_Dishoeck88,Bea03,Dutrey_ea07}.
The dust grains are assumed to be uniform $0.1\,\mu$m spherical particles made of amorphous silicates with olivine
stoichiometry \citep{Semenov_ea03}, with a dust-to-gas mass ratio of $1\%$.
The self-and mutual-shielding of CO and H$_2$ against UV photodissociation is computed using the pre-calculated
factors from Tables~10 and 11 in \citet[][]{Lee_ea96}.

Three other high-energy sources in the model that drive chemistry
either through dissociation, ionization, or desorption are cosmic
rays, decay of short-lived radionuclides, and stellar X-rays. The
X-ray ionization rate in a given disk region is computed according
to the results of
\citet{Glassgold_ea97a,Glassgold_ea97b} with parameters for their
high-metal depletion case. In this model, the
thermal $\sim 5$~keV X-ray photons are
generated at $\sim 0.1$~AU above the central star, with a total
X-ray luminosity of $\approx 10^{30}$~erg\,cm$^{-2}$\,s$^{-1}$
\citep{Glassgold_ea05}. The cosmic-ray ionization rate is assumed to
be 1.3$\times$10$^{-17}$~s$^{-1}$. The ionization rate caused by
radionuclide decay (primarily $^{26}$Al and $^{60}$Fe) is $6.5
\times 10^{-19}$~s$^{-1}$ \citep{Finocchi_Gail97}. We adding the
X-ray and radionuclide ionization rates to those involving cosmic
ray particles (CRP) and CRP-induced UV photons.

A modified static gas-grain chemical model that includes gas-grain
interactions without surface reactions and turbulent mixing has been
utilized \citep[][]{Semenov_ea05}. We allow only surface formation
of molecular hydrogen using the approach of
\citet{Hollenbach_McKee79}. This modification significantly
decreases the numerical demands of our chemical code and makes it
computationally tractable while allows us to include all necessary
physical processes. The gas-phase reaction rates and their
uncertainties are taken from the recent RATE\,06 database, in which
the effects of dipole-enhanced ion-neutral rates are included
\citep{rate06}. To calculate photoreaction rates through the disk,
we adopt pre-computed fits of \citet{van_Dishoeck88} instead of
integrating the wavelength-dependent cross sections over the local
UV spectrum \citep[see][]{van_Zadelhoff_ea03}.

Gas-grain interactions include the accretion of neutral molecules onto dust surfaces with a sticking
efficiency of 100$\%$, dissociative recombination of ions on charged dust grains, and grain re-charging, as well as
UV-, CRP-induced and thermal desorption of surface species.
Desorption energies $E_{\rm des}$ are mostly taken from the recent
osu.2007\footnote{\url{https://www.physics.ohio-state.edu/~eric/research.html}}
database \citep{GH_06} or roughly estimated by analogy for about 20 molecules.

Overall, our network consists of positively and
negatively as well as neutral dust grains, 420 gas-phase and 157
surface species made of 13 elements, and 5773 reactions. Among these
5773 reactions are 4517 gas-phase reactions, 2 charge exchange
reactions for dust grains, 940 dissociative recombination reactions
of ions on charged and neutral grains, and 314 accretion/desorption
processes.

As initial abundances, we have adopted the so-called ``low
metal'' set of \citet{Lee_ea98} as listed in Table~\ref{init_abund} and assumed that all hydrogen is mostly locked
in its molecular form. In these initial abundances,  the standard solar elemental composition is depleted
in heavy elements by, e.g., 200 times for S up to more than $10^4$ for Fe, Cl, P, and F.
The mostly atomic initial abundances are chosen instead of those from a molecular cloud model because this choice allows
the abundance uncertainties to accumulate in a long sequence of chemical reactions with imprecise rates, starting
from basic processes that reach steady-state at early times during the disk evolution. \citet{Willacy_ea98}
have shown that most of the molecular abundances are hardly affected by the choice of input abundances.
Note also that the resulting abundance uncertainties can be sensitive to the initial elemental concentrations
\citep{Wakelam_ea06b}. However, we do not allow the elemental concentrations to vary in order to keep the computations
in reasonable limits.

\subsection{Method to model rate uncertainties}
\label{method}
Each reaction rate coefficient in the RATE\,06
database is given in the standard Arrhenius form and thus relies on
3 parameters: $\alpha$ (the absolute value at room temperature),
$\beta$ (the index for the power law dependence of the rate on
temperature), and $\gamma$ (the activation energy barrier in K). A
specific expression for the reaction rate coefficient depends on the
type of chemical process: bimolecular reaction, direct cosmic-ray
ionization reaction, photodissociation, etc. Note that in all cases
the rate coefficient scales linearly with the parameter $\alpha$.

In addition to these three parameters, the RATE\,06 reaction rates are characterized by an
accuracy estimate (A, B, C, and D), where the uncertainties are smaller than $25\%$, $50\%$,
within a factor of 2, and within an order of magnitude.
Almost all reactions with measured rates belong to the first group
($\sim 1\,400$ reactions), while those reactions with rates that
were ``guessed'' or derived by analogy are assigned to the third
group ($\sim 2\,800$ reactions), which contains most of the
ion-neutral reaction rates. The second group includes about 300
mostly neutral-neutral and photodissociation reactions, whereas the
fourth group consists of only 4 photodissociation reactions
\citep{rate06}.

In accord with the latest osu.2007 network, we adopted a rate uncertainty of one order of magnitude
for radiative association reactions.
Such a large uncertainty factor is justified by the difficulty
in calculating or measuring these rate coefficients \citep[e.g.,][]{Herbst_85}.
The uncertainty factors vary widely for the photoreaction rates since the corresponding
frequency-dependent cross-sections have limited accuracy and have been
obtained only for a fraction of molecules in the UMIST database
\citep[][]{van_Dishoeck88,van_Dishoeck_ea06}.

Our method to model abundance uncertainties is based on the computation of a large set
of chemical models, using identical physical and initial conditions and the same
chemical network, but with randomly varied rate coefficients within their uncertainty
limits.
We utilized the same method to introduce uncertainties in the rate
coefficient values as described in
\citet{Dobrijevic_Parisot98}, \citet{Dobrijevic_ea03}, and \citet{Wakelam_ea05}. The rate
coefficient for each gas-phase reaction $i$ is randomly
chosen over a log-normal distribution with  median $\alpha_i$ and
dispersion $F_i$. Consequently, the rate coefficient of the $i$-th
reaction spans an interval between $\alpha_i/F_i$ and $\alpha_i\times
F_i$ with a probability of 68$\%$ ($1\sigma$). We generated a sequence
of these log-normally distributed rates for all gas-phase reactions
in our chemical network using the following expression:
\begin{equation}
    \alpha_i^l=\alpha_i\times(F_i)^{\epsilon},
\end{equation}
where $\alpha_i^l$ is the $l$-th realization of the $i$-th reaction
rate, $\alpha_i$ is the standard RATE\,06 rate value for the $i$-th
reaction, $F_i$ is the dispersion, or the uncertainty factor, for
this rate, and $\epsilon$ is randomly distributed from -1 to 1
with an uniform distribution law.

Using this approach, we allowed the rates of only 4517 gas-phase reactions
to vary in our network. The rates of all other processes (gas-grain interactions,
dissociative recombination of molecular ions on grains, and recharging of dust grains)
were kept constant.
Not did we account for possible variations among the parameters of the disk
physical model. Such an idealization allowed us to focus solely on an investigation of how  uncertainties in
gas-phase reactions affect the results of the disk chemical modeling.

Using 8\,000 realizations of the RATE\,06 network, we simulated 5~Myr of chemical evolution in the outer,
$r\ge 50$~AU, disk of DM Tau.
The disk grid consists of 5 radial (50, 100, 200, 380, and 760~AU) and 10 equidistant vertical points
(with step sizes of 3.2, 7.6, 18, 41.7, and 96 AU for the considered radii).
Such a huge number of utilized chemical models with randomly varied rates leads to low values of
statistical noise in the distributions of the abundance scatters at any particular disk location,
assuring the correctness of the sensitivity analysis. Moreover, a large number of varied reaction rates allows
an analysis of all reactions in the chemical network, in contrast to models with limited chemistry
included in previous works \citep[e.g.,][]{Wakelam_ea05}.  Still, even with the modest $5\times10$ spatial
resolution of the adopted disk model and without surface reactions, the overall computational time needed
to calculate the chemistry with these 8\,000 networks was about 3 days on a 4-CPU PC machine
(3.0~GHz Xeon, 8~Gb RAM).

\section{Results}
\label{res}

\subsection{Abundance distribution profiles}
\label{db_abunds}
Before we perform a detailed analysis of the influence of the rate uncertainties
on the modeled molecular abundances, we investigate the abundance distribution profiles.
Because we utilized a log-normal distribution
for the rate coefficients, one might expect that for each molecule the abundance
distribution should show a normal (Gaussian) profile in logarithmic scale.
If this hypothesis is correct, one can use the dispersion -- a measure of the scatter
of randomly varied values that have a Gaussian distribution -- as a convenient
quantity characterizing the abundance and column density uncertainties.

As shown in Paper~II, this assumption may not be fullfiled for certain species and certain time steps.
For example, when molecular abundances of a particular species show a steep decline or increase
with time then at that moment the corresponding histogram of
the abundance distribution can have several peaks and thus be far from a Gaussian shape because of sampling effects.
Bimodal distributions can also be obtained if the system is highly sensitive to a small variation of specific parameter.
\citet{Wakelam_ea06b} showed that dense cloud chemistry  hypersensitive, and even bistable, to
the ratio between He and H$_2$ cosmic-ray ionization rates in particular conditions.
In this case our uncertainty method would tend to overestimate the dispersion of the modeled abundances.
The disk chemistry is very rich in a sense that it proceeds vastly differently in various disk
regions and at different times, so it is natural to expect that the abundance distributions
of some species can sometime deviate from a normal distribution.

We carefully studied this problem and found that such a situation happens rarely at late times,
$\ga 1$~Myr, for which our analysis will be performed. In Fig.~\ref{hists}
we show several representative histograms of the abundance distributions at 5~Myr, using
HCO$^+$ and CO as an example. For an inner disk region, at $r\approx 97.5$~AU, $z \approx 30$~AU,
a strong deviation from the Gaussian shape is clearly visible, with 2 distinct peaks and a gap
in between (Fig.~\ref{hists}, top panels). Note that this is one of the few exceptional cases
that is discussed in detail
below. Everywhere else through the disk the HCO$^+$ and CO abundances have nearly Gaussian-like profiles
(Fig.~\ref{hists}, bottom panels). This condition holds for other key species as well, implying that the
averaged abundances in the disk should not differ from the values computed with the standard RATE\,06 network.
Moreover, in Paper~I we found that the Gaussian shape of the abundance distribution
is also preserved for the uniform distribution of the rate uncertainties in linear scale.

In the rest of this paper,  we will refer to  one standard deviation ($1\sigma$) in
the modeled abundance distribution as  the ``abundance uncertainty''. In our notation ``abundance scatter''
represents twice the abundance uncertainty in logarithmic scale ($2 \log \sigma$).
It is the abundance uncertainties that have to be taken into account
when interpreting the observational data with a chemical model.

\subsection{Distribution of the mean abundances in the disk}
\label{mean_abunds}
Although the evolution of all species was calculated in  8\,000 realizations of the chemical model, in what
follows we focus on several key molecules that are used as tracers of disk physical structure and
chemical composition.
Our list of important species includes molecules that are widely used for kinematic studies and
determination of the density and temperature (CO, CS, H$_2$CO, NH$_3$, and HCN)
as well as for probing such major parameters as ionization degree (HCO$^+$),
radiation fields (CN, HCN), and the yet-to-be detected C$^+$ as a dominant ion in the disk surface
\citep{Dutrey_ea97,Kastner_etal1997,Aikawa_ea03,Dutrey_ea07}.

We use time-dependent abundances to analyze the chemical evolution of these species in detail.
The analysis is based on the calculation of the relative importance of each reaction in the chemical network
for the abundance change of a molecule under investigation at a given time \citep{Wiebe_ea03}.
This allows us to weight all reactions in the model and isolate a few most important pathways
for chemical evolution of one or several species under specific physical conditions.
In the case of protoplanetary disks, such an analysis has to be performed for several tens of representative cells.

In Fig.~\ref{Abu2D} we show the mean abundances of these molecules
in the disk at 5~Myr, which are obtained by averaging the results of
the chemical network. The vertical extension of the disk is scaled
by one hydrostatic scale height, $H(r) = \sqrt{2}C_s/\Omega$, where
$C_s$ is the midplane sound speed and $\Omega$ is the Keplerian
frequency \citep{Dartois_ea03}. The unity pressure scale height is
about 25 and 320~AU at distances of 100 and 800~AU, respectively.

The ``sandwich''-like chemical structure of the disk is clearly seen in Fig.~\ref{Abu2D}.
In the ``dark'', cold, and dense midplane at  $\la 0.5-1$ pressure scale height,  the molecules
freeze out onto dust grains within $100-100\,000$~yr and low steady-state fractional abundances of
$\sim 10^{-14}-10^{-10}$are reached. The abundances of the gas-phase species
in the midplane at late times are sustained by the CRP-induced desorption of the mantle materials.
The midplane abundance of CO $\sim 10^{-8}$, is higher because this molecule is very abundant
($\approx 10^{-4}$) in the gas phase initially, and since CO ice is rather volatile because of its low
CO desorption barrier of $\sim 930$~K \citep{Bisschop_etal2006}.
Overall, the gas-phase abundances in the midplane tend to slightly increase with radius as the disk
surface density and thus the density in midplane decrease and gas-grain interactions become less intense,
which slows down the freeze-out of gas-phase molecules onto dust grain surfaces.

The surface layer of the disk ($Z/H(r)\ga 2$) is deficient in molecules due to
strong irradiation by high-energy stellar and interstellar photons. Even CO, which is self-shielded and
mutually-shielded by H$_2$ against UV photodissociation cannot survive. Among the considered species, only C$^+$
reaches its maximal concentration and becomes the most abundant ion in the surface layer, with an abundance of
$\sim 10^{-4}$. The thickness of this highly ionized layer ($\sim 1$ pressure scale
height down from the disk surface) stays nearly constant with radius for all key species but highly reactive
radicals CS and CN, whose abundances are already high ($\sim 10^{-9}$) at scale pressure heights as high as
$Z/H(r) \approx 2$ at all radii (Fig.~\ref{Abu2D}).

In contrast, polyatomic species tend to concentrate in the warm, chemically rich intermediate layer
($\la 1-2$ pressure scale height, depending on the radius). The model-averaged
abundances for the considered species, $\sim 10^{-10}-10^{-4}$, are close to the values computed
with the standard RATE\,06 network and those found in previous work
\citep{Aikawa_Herbst99,Willacy_Langer00,Aikawa_ea02,Willacy_ea06,Dutrey_ea07}.

Now let us consider the chemical processes that form and destroy the considered species in the intermediate
layer. In contrast to dense cloud models, for many species in this disk zone steady-state is
reached by $\sim 10^3-10^5$ yr. This is due to the higher densities encountered in the warm layer
($\ga 10^6-10^8$~cm$^{-3}$)
than in the cloud cores ($\sim 10^4-10^6$~cm$^{-3}$) and a non-negligible flux of the UV radiation. Impinging interstellar
UV photons allow many surface species to desorb back into the gas phase \citep{Hasegawa_Herbst93,GH_06},
thus reducing their surface abundances and overall importance of gas-grain interactions for
disk chemical evolution at late times, $t\ga 10\,000$ yr.

The evolution of CO and HCO$^+$ in the intermediate layer is
governed by a small set of reactions at all radii. The CO molecules
are formed via reactions of atomic oxygen with CH, CH$_2$, and C$_2$
at $t\la 1\,000$~yr, and partly through the dissociative
recombination of HCO$^+$ afterwards (at this stage steady-state has
been reached and there is a loop between CO and its protonated
form). The light hydrocarbons that are precursors of CO are
themselves rapidly created within a few years via the radiative
association of C with H and H$_2$, and CH with H$_2$. The CO is
mainly removed from the gas phase by accretion onto the dust grain
surfaces in outer, cold regions and by reactive collisions with
helium ions in the inner, X-ray ionized part of the intermediate
layer. In the upper part of this layer, CO is also photodissociated.
The ion HCO$^+$ initially forms via reactions of O with CH and
ionized light hydrocarbons, CH$^+_{\rm n}$ (n=1--3), and the
ion-neutral reaction between CO and CH$_5^+$ (also producing
methane). At a later time of $\ga 1\,000$~yr, HCO$^+$ becomes the
most abundant ion in the intermediate layer and the dominant molecular
ion in the disk, with a fractional abundance of
$\sim 10^{-9}$. At this steady-state stage, HCO$^+$ is mainly
produced via reactive collisions between CO and H$_3^+$ and
destroyed by  dissociative recombination with electrons and
negatively charged grains.

The chemically related H$_2$CO molecule is produced through reaction
of CH$_3$ with oxygen atoms. At a later time of $\sim10^5$~yr
H$_2$CO is involved in a simple formation-destruction cycle. It
starts with protonation of formaldehyde by HCO$^+$, H$_3$O$^+$, and
H$_3^+$, which is followed by dissociative recombination into
H$_2$CO (33\%), CO (33\%), or HCO (33\%). Apart from
photodissociation and freeze out, another direct destruction channel
for H$_2$CO in the upper, more ionized part of the intermediate
layer is the ion-neutral reaction with C$^+$, which
produces either CO and CH$_2^+$, HCO$^+$ and CH, or H$_2$CO$^+$ and C.
The chemical evolution of formaldehyde typically reaches steady-state at about $10^5$~yr
with a gas-phase abundance of $\la 3\times10^{-10}$.
Note that in the same disk chemical model as discussed above but with a set of surface
reactions taken from \citet{GH_06} the overall (gas-phase and surface) abundance
of formaldehyde is decreased by a factor of several as it is converted to methanol by hydrogenation
on grain surfaces.

The photostable radical CN is abundant ($10^{-8}$) at an upper, less molecularly rich disk layer
($Z/H(r) \approx 1.8$, Fig.~\ref{Abu2D}). The initial formation channel for the cyanogen radical is the radiative
association between C and N, followed by the neutral-neutral reaction of N and CH, and dissociative recombination
of HCNH$^+$ at later times ($t\ga 10^4$~yr).
Other, less important formation pathways for CN at this stage are neutral-neutral reactions between
C and NO producing CN and O as well as between C and OCN leading to CN and CO.
The destruction of CN mostly proceeds through photodissociation and neutral-neutral reaction
with N forming highly photostable nitrogen molecules and C.

In turn, HCNH$^+$ is produced by ion-neutral reactions
between H$_2$ and HCN$^+$ and between N and C$_n$H$_2^+$ ($n=3,4$)
at early times ($t\la 10^3$~yr). Later the steady-state recycling of
HCNH$^+$ is reached, which involves dissociative recombination of
HCNH$^+$ into either HCN (33\%), HNC (33\%), or CN (33\%), followed
by re-production of HCNH$^+$ via protonation of HCN and HNC by
either H$_3^+$, H$_3$O$^+$, or HCO$^+$ as well as via the
ion-neutral reaction of CN with H$_3^+$.

The chemistry of HCN is tightly related to the evolution of CN and
HCNH$^+$. HCN is mainly produced by neutral-neutral reactions
involving nitrogen and methylene or methyl at early evolutionary
stages, $t\la 10^4$~yr. Later the evolution of HCN proceeds as a
part of the HCNH$^+$ formation-destruction loop. The major
destruction routes for  gas-phase HCN include freeze-out onto dust
grains and the ion-neutral reaction with C$^+$, which forms
CNC$^+$ and H. Another important destruction channel for HCN in the
upper disk layers at $\sim 1.5-2$ scale heights is
photodissociation. Steady-state for HCN is typically reached within
$\ga 10^4$~yr, with an abundance of about a few $\times 10^{-10}$.

The layer where abundant ammonia exists is located somewhat deeper
toward the disk midplane compared with HCN and especially CN because
NH$_3$ can be photodissociated by UV photons shortward of $\approx
1\,950$~\AA \citep{van_Dishoeck88}. The main formation route for
ammonia is dissociative recombination of NH$_4^+$, which itself is
formed through a sequence of hydrogen insertion reactions starting
with the production of the NH$^+$ ions from N$^+$ reacting with
H$_2$. Upon formation, NH$^+$ repeatedly reacts with molecular
hydrogen and gains an addition hydrogen atom until NH$_4^+$ is
formed. The destruction of ammonia proceeds mainly via
ion-neutral reactions with C$^+$ and S$^+$ in the upper
intermediate layer as well as H$_3^+$, H$_3$O$^+$, and HCO$^+$ in a
deeper region, which leads again to NH$_4^+$ and either H$_2$,
H$_2$O, or CO. For all radii, the abundance of ammonia reaches a
steady-state value of $\la 3\times 10^{-10}-10^{-9}$ at very late
times of $\ga 10^5-10^6$~yr.

The chemical evolution of CS is governed by neutral-neutral
reactions between sulphur atoms and either CH, CH$_2$, or C$_2$
initially, and later, at $t\ga 10^3$~yr, by an extended
formation-destruction cycle of H$_{\rm n}$CS$^+$ (n=1--3) and
HC$_2$S$^+$. At this evolutionary stage these complex ions are
produced by ion-neutral reactions of protonated water,
H$_3^+$, and HCO$^+$ with H$_{\rm n}$CS (n=0--2). The initial routes
to form the HCS$^+$ ion are the reactions between CH$_3^+$ and S or
CS$^+$ and H$_2$, whereas HC$_2$S$^+$ forms predominantly via
reaction of CH$_3$ and S$^+$. The H$_3$CS$^+$ ion is mostly produced
in reactive collisions of methane with sulphur ions. Upon formation,
HCS$^+$ dissociatively recombines into CS ($\sim 10\%$) or CH ($\sim
90\%$), while HC$_2$S$^+$  breaks up equally into C$_2$S or CS. The
H$_2$CS$^+$ and H$_3$CS$^+$ ions recombine with equal probabilities
into CS and either HCS or H$_2$CS. Finally, the major destruction
channels for CS are photodissociation and the ion-neutral
reaction with C$^+$. The steady-state abundance for CS, $\sim
10^{-9}$, is usually attained at about $10^3-10^5$~yr in our model.

\subsection{Distribution of the abundance and column density uncertainties}
\label{err_dist}
For each key molecule, we derived the distribution of abundance uncertainties (dispersion) in the disk.
The results are shown in Fig.~\ref{UncertDistrib}. These uncertainties are quite different in different disk regions.
Note that the overall abundance uncertainties are not much larger for the more complex (heavy) species
among considered, in comparison to those with fewer atoms (e.g., compare C$^+$, CO, HCO$^+$, and H$_2$CO).
However, if we would focus on even more complex species (e.g., consisting of $>7$ atoms), the abundance uncertainties
will in general increase with the number of atoms in the molecule, as shown by \citet{Vasyunin_ea04} and \citet{Wakelam_ea06}.

Among the key reactions, the rates of photodissociation and
photoionization processes can be uncertain by factors of 2 and
larger. Moreover, their absolute values depend on the details of the UV radiative transfer
modeling \citep{van_Zadelhoff_ea03}.
Thus, it is not surprising that in the disk
atmosphere at $Z/H(r)\ga 1.5-2$ large abundance uncertainties with a
factor of $\la 10-30$ are reached for most of the species, in
particular across the interface between the intermediate layer and
the disk atmosphere. A notable exception is ionized carbon, because
it contains essentially all of the element carbon in the disk
atmosphere and thus maintains a high steady-state abundance that is
not affected by chemistry. This effect leads to relatively low
values of the C$^+$ abundance uncertainties, in particular a factor
of $\la 3$. The same is true for CO,  which locks up all carbon in
the disk midplane and intermediate layer at $\la 2$ pressure scale
height and here has its lowest abundance uncertainty, a factor of
$\la 2$ (Fig.~\ref{UncertDistrib}).

There is a disk region adjacent to the midplane at $r\sim 100$~AU, $Z/H(r) \sim 1$, where
the abundance uncertainties of the considered carbon-bearing species including CO show an increase of up to one order
of magnitude (see HCO$^+$, Fig.~\ref{UncertDistrib}).
There, the chemistry of CO and hence all chemically related species
(e.g., HCO$^+$, H$_2$CO, etc.) are prone to the hypersensitivity at $\ga 10^6$~yr
caused by the rate uncertainty of the X-ray ionization of helium atoms, as
discussed in \citet{Wakelam_ea06b}.
This region is moderately warm, with a temperature $T\sim 25-40$~K, and well shielded from the UV radiation
from the central star and the interstellar
UV radiation ($A_{\mathrm V} \la 2-5$~mag). The thermal bremsstrahlung X-ray photons ($\sim 5$~keV)
generated at
$\sim 0.1$~AU above the star, however, are able to penetrate into this region \citep{Glassgold_ea97a},
with a total ionization rate of $\sim 10^{-16}$~s$^{-1}$. These X-ray photons produce helium
ions that rapidly react and destroy the CO molecules (forming O and C$^+$). In turn, these C$^+$ ions are used
to reform CO but also to slowly produce carbon chain molecules (e.g., C$_2$H$_2$) and cyanopolyynes (e.g., HC$_3$N),
which are removed from the gas phase by freeze out onto the dust surfaces. These heavy surface molecules lock up most of
the elemental carbon at $t\ga 1$~Myr such that the gas-phase abundance of CO is reduced to $\sim 10^{-9}-10^{-7}$.
The overall efficiency of the carbon chain formation and the decline of the CO abundance at late times
sensitively depend on the He ionization rate. Consequently, the CO histogram of the abundance distribution at 5~Myr
has a double-peaked, non-Gaussian shape, as shown in Fig.~\ref{hists}.

The abundances of nitrogen-bearing species such as CN, HCN, and NH$_3$ in the region around the midplane at
$r>200$~AU (Fig.~\ref{UncertDistrib}) possesses a high sensitivity to the rate of the CRP ionization of He.
The histograms of CN, HCN, and NH$_3$ are not fully Gaussian at 5~Myr in this region, and our method tends
to overestimate the resulting abundance uncertainties.
This region is so cold that at late times, $t\ga 3\times 10^5$~yr,
most molecules are removed from the gas phase by freeze-out. At this
evolutionary stage, H$_3^+$ and later H$^+$ become the most abundant
charged species. The primal formation pathway for H$^+$ is the slow
ion-neutral reaction between He$^+$ and H$_2$. The rise of the H$^+$
abundance at $\sim 5$~Myr($ \la 10^{-8}$) occurs in part because of
the slowness of the radiative recombination of ionized hydrogen
atoms in the molecular-deficient gas. The total electron
concentration increases in tandem with H$^+$, and leads to more
rapid dissociative recombination of HCNH$^+$ and NH$_4^+$ (precursor
ions for CN, HCN, and NH$_3$). Consequently, the previously low
gas-phase abundances of these nitrogen-bearing molecules increase at
$t\sim5$~Myr by factors of several. The steepness and evolutionary
pattern for these abundance gradients depend sensitively on the
abundance of helium ions.

In the rest of the intermediate layer,  the chemical evolution of
the gas-phase species is initiated by radiative association
reactions with high uncertainties but governed later by
ion-neutral and neutral-neutral reactions with many accurately
estimated rates. The situation in the disk midplane is similar,
though at later times the chemical evolution of many gas-phase
species is governed by the steady-state accretion-desorption
life-cycle, which has no uncertainties in our model. This effect
leads to lower abundance uncertainties of a factor of $\sim 3$ than
occur in the upper disk region (see Fig.~\ref{UncertDistrib}).

It is interesting to note that for CO, C$^+$, and NH$_3$ the uncertainty peak does not coincide
with the maximal relative abundance in the disk. One might expect that the scatter in their column
densities could be smaller than for other considered species (CS, CN, HCN, HCO$^+$, and H$_2$CO),
for which uncertainty peaks and maximum abundances tend to coincide.

In Fig.~\ref{col_dens}, the radial distributions of the column
densities at 5~Myr are plotted for the considered species (dotted
lines). In addition, in Table~\ref{abs_uncer} we compile typical
intrinsic uncertainties of the column densities of a larger set of
key molecules in protoplanetary disks at 100 AU.
Indeed, the overall uncertainty in the column densities of CO,
C$^+$, and NH$_3$ as well as H$_3^+$, H$_2$O, N$_2$H$^+$, and
HCNH$^+$ is less than a factor of about 2. Using this uncertainty
value as a criterion, we assign these species to the so-called first
uncertainty group. On the other hand,  the column densities of CS,
CN, HCN, HCO$^+$, and H$_2$CO as well as CO$_2$, C$_2$H, and HNC
have larger error bars up to a factor of 4. These species belong to
the second uncertainty group. It is likely that the species from the
first sensitivity group are more reliable observational tracers than
the species from the second one.

Note that the column density uncertainties do not exceed a factor of about 4 even for formaldehyde
and are comparable with observational uncertainties. However, for heavier species (e.g., HC$_9$N),
which are not included in our study, the uncertainties are likely larger, as shown in Paper~I and Paper~II.
In our model, these uncertainties slowly
increase with time and typically reach steady-state at $\ga 10^5$~yr.
The column density uncertainties can be used as error bars of the theoretically
predicted quantities when comparing with observational data.
Moreover, one can envisage a situation when, for an object with a well-studied structure, high-resolution
observations followed by advanced modeling will allow putting such tight constraints on molecular abundances or
column densities that it may help to determine the rate coefficients of some key reactions \emph{observationally}.

\section{Discussion}
\label{diss}
In the previous section, we demonstrated how and to what extent the uncertainties of the rate
coefficients affect the computed abundances and column densities
of several key species in the disk. In this section, we discuss the relative
contributions of individual reactions to the uncertainties of these key molecules.

\subsection{Sensitivity analysis method}
\label{corr_method}
The chemical evolution of even a simple molecule usually involves a large number of reactions in the adopted
chemical network. The non-linearity of the equations of chemical
kinetics often makes it a challenge to find a direct link
between the rate value of a certain reaction and the molecular
abundances. Thus an efficient numerical method is required \citep[see, e.g.,][]{Turanyi_97}.

Using a sensitivity analysis based on a simple correlation method, we have shown in Paper~I
that under the physical conditions of dark and diffuse clouds the rate uncertainties of only a handful of
chemical reactions affect the accuracy of the resulting abundances to the highest extent.
For the disk chemistry studied in this paper a modified, two-stage sensitivity analysis was performed.

First, for each selected molecule, we calculated time-dependent
linear correlation coefficients between the abundances and rate
coefficients for all 8\,000 realizations of the RATE\,06 network,
and 19 logarithmically taken time steps over the
5~Myr of disk evolution at each disk cell. The coefficients
$R_{ij}(t)$ are given by the expression
\begin{equation}
R_{ij}(r,z,t)=\frac{\sum_{l}{(x_i^l(r,z,t)-\overline{x_i(r,z,t)})(\alpha_j^l-\overline{\alpha_j})}}{
                \sqrt{\sum_{l}{(x_i^l(r,z,t)-\overline{x_i(r,z,t)})}^2\sum_{l}{(\alpha_j^l-\overline{\alpha_j})}^2}},
\end{equation}
where $x_i^l(r,z,t)$ is the molecular abundance of the $i$-th species
computed with the $l$-th realization of the UMIST network,
$\overline{x_i(r,z,t)}$ is the standard (mean) abundance of this species,
$\alpha_j^l$ is the $j$-th reaction rate coefficient for the $l$-th
realization of the RATE\,06 network, and $\overline{\alpha^j}$ is
the standard rate coefficient of the $j$-th reaction.

This is in contrast to Paper~I where the
analysis of statistically significant reactions was performed for only the final time step.
Since at various evolutionary stages the chemistry of a molecule is typically governed by a restricted number
of chemical processes, at any particular time step the abundance correlation will be high only for some of the
key reactions, as shown in Fig.~\ref{CorrInTime} for the case of ammonia.
The obvious drawback of such a ``single-time'' approach is that it does not
account for the cumulative character of the abundance uncertainties.
Therefore, we integrated the absolute values of
time-dependent correlation coefficients over the entire evolutionary
time of 5~Myr. These quantities are called ``cumulative correlation
coefficients''.

With this cumulative criterion, one can identify those reactions whose rate uncertainties are the most
important for the abundance scatter of one molecule in one specific disk region. As a general criterion for
the entire disk, we utilized the cumulative correlation summed up over all 50 disk grid cells, followed
by summation over the 8 considered species (CO, C$^+$, CS, HCO$^+$, H$_2$CO, CN, HCN, NH$_3$).
One should bear in mind that the results of our analysis
rely on the inherent assumption that the absolute values of the rate coefficients and their uncertainties
in RATE\,06 are accurately determined. This assumption may not be fullfiled for all important reactions in the
RATE\,06 database as we will show below. Such reactions will not be treated correctly by our sensitivity analysis.

\subsection{Identification of the most problematic reactions}
\label{cons_impr}
We calculated disk-averaged cumulative correlations for all 5773 reactions in our chemical network and for
8 key species: CO, C$^+$, CS, HCO$^+$, H$_2$CO, CN, HCN, and NH$_3$. The corresponding cumulative correlations,
normalized in descending order, steadily decline with the number of reaction. The 56 most ``problematic''
reactions and their weights for these 8 molecules are listed in Table~\ref{imp_reac}. We term by ``weight''
the corresponding cumulative correlations summed over the all 50 disk cells and renormalized to the maximum value
of $1148.1$, which is achieved for the RA reaction between H$_2$ and C.

These 56 reactions correspond to
$1\%$ of the total number of reactions in our chemical model. Moreover, such an apparently small number of reactions
is chosen to stimulate experimental and theoretical studies of the rate coefficients of the reactions with high
astrochemical importance. Note also that some of the most ``problematic'' reactions in Table~\ref{imp_reac} have
been carefully studied as their rates have small uncertainties of only factors of 1.25--1.5.

In order to demonstrate that these 56 reactions are indeed important and their rate uncertainties
strongly affect the abundances, we re-calculated the disk chemical evolution but allowed the rates of these 56
reactions to vary by only a factor of 1.25 (and 2 for radiative association). Such idealization is thought to
mimic the situation when all of these 56 rates will be accurately measured in the laboratory or derived theoretically.
%

The resulting uncertainties in the molecular column densities are shown in Fig.~\ref{col_dens} (solid lines)
and reported in {Table~\ref{imp_reac} (last column). Compared with the initial column density uncertainties,
the refined chemical model leads to smaller error bars for most of the considered molecules, in particular
for CO, HCO$^+$, H$_2$CO, CN, and H$_2$O. However, the improvement is not so obvious for nitrogen-bearing species
(HCN, HNC, NH$_3$, HCNH$^+$) as well as CO$_2$, CS, and C$_2$H. Their chemical evolution is governed by a large set
of key reactions that are only partly included in the list of the 56 most problematic reactions
(Table~\ref{abs_uncer})\footnote{An extended version of this table is freely
available from the authors upon request.}.

The fact that the uncertainties in column densities decrease not only for the 8 key species, but also for
some other molecules in Table~\ref{abs_uncer} clearly indicates that some of the reactions from Table~\ref{imp_reac}
are relevant for their chemical evolution. For H$_3^+$ these are the ionization of H$_2$ and reactions of H$_3^+$
with C, O, and CO. As we have shown in Sect.~\ref{db_abunds}, reactions with H$_3^+$ are an essential ingredient
of disk chemistry. The decrease in the abundance error bars for H$_3^+$ in the model with refined rate
uncertainties leads to smaller abundance uncertainties in the related species: HCO$^+$, H$_2$CO, H$_2$O,
HCN, HNC, HCNH$^+$, and CS.
Table~\ref{imp_reac} contains a number of reactions important for the evolution of CO and CO$_2$;
e.g., formation of light hydrocarbons,
neutral-neutral reactions of these hydrocarbons with oxygen, and CO photodissociation. The former reactions
affect the evolution of C$_2$H, while photodissociation is an important process for CN.

Below we discuss the contributions of each reaction type in detail.

\subsubsection{Radiative association}
\label{RA_imp} The most problematic reactions in the disk are
radiative association (RA) reactions. Radiative association leads to
the formation of a larger molecule from two smaller species upon
their collision, and requires  the emission of excess energy in the
form of a photon. RA reactions allow the formation of new bonds and
more chemically ``advanced'' species \citep{van_Dishoeck_Blake98}.
The corresponding rate values are often so low for small species
that they are difficult to obtain with high accuracy
\citep{Williams72,Herbst_80,Bates_Herbst88a,Smith89}. Nonetheless,
assigned uncertainties for many of these reactions in RATE\,06 are
only a factor of 2 and less. Since we assume that the rate
coefficients of such reactions may vary by an order of magnitude, in
accord with the recent OSU database, it is not surprising that some
RA reactions are identified as among the most problematic reactions
for disk chemistry. Its importance is also rise due
to the fact that we start calculation from the atomic initial
abundances and use time-integrated correlation coefficients. It
means that at early evolutionary time chemistry is dominated by RA
reactions and, therefore, its time-integrated correlation
coefficients increase greatly at early times.

Two RA reactions are particularly important for the chemical evolution and accumulation of the abundance uncertainties.
First is the reaction between H$_2$ and C, which produces CH$_2$ everywhere through the disk at all times.  The rate coefficient of this reaction is not well known, with
with an estimated value of $10^{-17}$~cm$^3$\,s$^{-1}$ at $T = 10-300$~K \citep{Prasad_Huntress80,rate06}.
This reaction is relevant for the evolution of CO, HCO$^+$, H$_2$CO, CS, and HCN. 
Second, the reaction between H$_2$ and C$^+$ leads to the formation of
CH$_2^+$ (important for C$^+$, HCO$^+$, CS, NH$_3$, and CN),
with an estimated rate coefficient of $\approx 4\times10^{-17}$~cm$^3$\,s$^{-1}$ \citep{Herbst_85}. These
two reactions initiate processes of C and H addition in the chemical
model and thus should be relevant for nearly all species containing
carbon and hydrogen.

The next most important RA reaction (8th place out of 56) is the formation of protonated methane from molecular hydrogen
and CH$_3^+$ \citep[$\alpha=1.3\times 10^{-14}$~cm$^3$\,s$^{-1}$, $\beta=-1$;][]{Smith89},
which is an important molecular ion for CO destruction.
More recently the rate of this reaction has been measured at low temperature of 10~K
by \citet[]{G95}, $1.1\pm0.1 \times 10^{-13}$~cm$^3$\,s$^{-1}$ for para-H$_2$.

Much less important (27th and 28th place) are two slow reactions
between C and H or N producing CH or CN radicals. It is interesting
to note that the latter reaction has different rate coefficients in the
OSU and RATE\,06 databases \citep{OSU03,rate06}. While in
osu.2007 this reaction rate has no temperature dependence and no activation
barrier \citep[$\alpha=10^{-17}$~cm$^3$\,s$^{-1}$;][]{Prasad_Huntress80}, in RATE\,06
the corresponding rate is low, $7.9\times10^{-19}$~cm$^3$\,s$^{-1}$, and possesses a small barrier of 96~K
\citep[relevant temperature range is between 300 and 2\,700~K;][]{SA00}.
Thus the RATE\,06 and osu.2007 rate coefficients for
the RA reaction between C and N differ by an order of magnitude in a wide temperature range.

The situation is similar for the RA reaction between molecular hydrogen and CH
that forms CH$_3$ -- an important radical for the evolution of formaldehyde.
The corresponding osu.2007 rate at 10~K is $2.5\times 10^{-16}$~cm$^3$\,s$^{-1}$
($\alpha=3.25\times 10^{-17}$~cm$^3$\,s$^{-1}$, $\beta=-0.6$). In contrast, in RATE\,06 this
reaction has a small barrier of 11.6~K and the 10~K rate value is lower by an order of magnitude,
$1.8\times 10^{-17}$~cm$^3$\,s$^{-1}$
\citep[$\alpha=5.09\times 10^{-18}$~cm$^3$\,s$^{-1}$, $\beta=-0.71$, and $\gamma = 11.6$~K;][]{BS97}.

In addition to radiative association, we list radiative recombination
reactions of C$^+$ and CH$_3^+$ in the same group in Table~\ref{imp_reac}.
The radiative recombination of ionized carbon is a key
reaction for the evolution of C$^+$. 
Its rate coefficient at room temperature is $4.67\times
10^{-12}$~cm$^3$\,s$^{-1}$ with an inverse temperature exponent of
0.6 \citep[][]{rate06}. The assumed uncertainty of this rate
(the factor of 10) in osu.2007 is in contrast to the results of
\citet{Nahar_Pradhan97}, who have theoretically predicted this rate
value and found that the corresponding uncertainty is only $\sim
10\%$. In the RATE\,06 database the uncertainty of
this rate constant is set to 50\%.

Finally, the radiative recombination of CH$_3^+$ is the
least probable channel in the reactive collisions between methyl
ions and electrons. The rate of this process is estimated to be much higher than for
other recombination reactions involving atomic species, particularly in cold
regions, and might be accurately determined in laboratory
\citep[$\alpha=1.1 \times 10^{-10}$~cm$^3$\,s$^{-1}$,
$\beta=-0.5$;][]{umist95,umist99,rate06}. In the osu.2007 database, this rate has a
steeper dependence on temperature, $\beta=-0.7$ \citep{OSU03}.

We conclude that RA reactions require a particular attention
because their rate uncertainties are often large and not well known so that the modeling results can be heavily affected.

\subsubsection{Cosmic ray ionization}
\label{CRP_imp}
The next most important reactions that contribute strongly to the resulting abundance scatter
of many molecules are the cosmic ray ionization of molecular hydrogen and helium atoms.
The cosmic ray particles able to penetrate deeply into the disk are mostly high-energy protons
($E\ga 100$~Mev) and heavy nuclei (e.g. iron, $E\sim 1$~Gev)
\citep{DalgarnoMacCray72,LJO85,DolginovStepinski94}.
Direct ionization initiate chemical evolution in dark
and cold regions where cosmic ray particles remain the main ionization source \citep{Herbst_Klemperer73}.
In the disk this is the outer midplane at $r\ga 30$~AU and the lower part of the intermediate
layer \citep{Gammie96,Semenov_ea04}. 
Moreover, since we added the X-ray ionization rate to the CRP rate value in an attempt to crudely treat the effect
of high-energy stellar radiation, it is not surprising that high correlation coefficients for
some CRP-ionization reactions are reached in the disk atmosphere and intermediate layer at $r\la 100$~AU
for several key species such as CO (see Fig.~\ref{UncertDistrib}).
A more accurate approach to model X-ray chemistry in disks is required
\citep{Neufeldea94,Maloneyea96,LeppDalgarno97,Staeuber_ea05}.

It is natural that because the cosmic ray ionization of H$_2$ and He is the primal chemical process, it is
also the key factor for the accumulation of abundance scatters.
In the RATE\,06 database the corresponding rate coefficients are estimated to be accurate within a factor of 2
\citep{rate06}. However, this rate uncertainty has two origins.

First, there is the \emph{physical} uncertainty that is related to our limited knowledge of
the absolute CRP flux and its penetrating ability, which may vary in different astrophysical objects
\citep{Spitzer_Tomasko68,DolginovStepinski94,Caselli_ea98}.
While the direct ionization process driven by CRPs is well studied and the corresponding cross sections and rates
are well known \citep{Solomon_Werner71}, the second \emph{chemical} uncertainty resides in the treatment of the
ionization by energetic secondary electrons \citep{Glassgold_Langer_73}. Here the elemental composition
of the gas plays a decisive role \citep{Ilgner_ea06a}.

The CRP-induced ionization rate of H$_2$ currently used in all UMIST
and OSU databases was taken from the PhD thesis of Black (1975). It
was computed for a mixture of atomic and molecular hydrogen but new
calculations for a mixture of hydrogen and helium would be required
in order to have a more consistent rate value \citep[see discussion
in][]{Wakelam_ea06b}. Other CRP-ionization reactions listed in
Table~\ref{imp_reac} do not have a reference in the RATE\,06
database so it is hard to judge how accurate their rates are.

Apparently, the rate uncertainty of the CRP-driven reactions can be partially eliminated
by more accurate modeling of the full cascade of ionizing events, but the physical uncertainty will
still remain, which needs more efforts from observers. \citet{Wakelam_ea06b} have demonstrated how crucial this
may be for the results of pure gas-phase chemical models of static molecular clouds.

\subsubsection{Photoreactions}
\label{PD_imp}
Another set of chemical processes with problematic rates are
photodissociation (PD) and photoionization (PI) reactions, as listed in Table~\ref{imp_reac}.
These reactions are important in those regions where UV photons are either able to penetrate
(the disk atmosphere and intermediate layer) or are induced by cosmic ray particles
\citep[disk midplane;][]{PrasadTarafdar83}.
High-energy UV photons destroy or ionize gas-phase molecules and allow surface species to desorb back into the gas
phase.

While both the PD and PI cross sections for many species can be
measured or calculated with a rather good accuracy of $\la 50\%$
\citep{van_Dishoeck88,van_Dishoeck_ea06}, the major difficulty in
obtaining accurate photorates is our limited knowledge of the UV
radiation field inside protoplanetary disks. \\citet{Bea03}
and \citet{van_Dishoeck_ea06} have shown that many T Tauri stars
emit strong, non-blackbody UV radiation. Their spectra resemble that
of the interstellar (IS) UV radiation
field, with a large fraction of the UV flux emitted in the
Ly$\alpha$ line. \citet{van_Zadelhoff_ea03} have demonstrated the
importance of scattering for the UV penetration deep into the disk
interior. Furthermore, for the two most abundant molecules, H$_2$
and CO, that effectively dissociate through partly overlapping lines
shortward of about $1\,120$~{\AA}, the optical thickness in these
lines can become so high that self- and mutual-shielding have to be
taken into account
\citep{DS70,van_Dishoeck88,Lee_ea96,Jonkheid_ea07}.

In the UMIST and OSU databases, the photorates have mostly been
adopted from the compilation of \citet{van_Dishoeck88} and
\citet{RJLD91}, where they have been computed using a simple
plane-parallel approach and the IS (IS) UV radiation
field of \citet{Draine_78}. The IS UV flux is characteristic of a
diluted radiation field from an early B star ($T_{\rm eff}\sim
30\,000$~K) located at a distance of $\sim10$~pc, with a cut-off at
912~{\AA} due to absorption by interstellar atomic hydrogen
\citep{Spitzer_78,vanDishoeck94}. Obviously, it is not
representative of the UV radiation from much cooler Herbig Ae and T
Tauri stars surrounded by circumstellar disks. Recently,
\citet{van_Dishoeck_ea06} have recalculated the corresponding PD and
PI rates, using more appropriate UV spectral distributions typical
of Herbig Ae, T Tauri, and Sun-like stars.

The 10 problematic photodissociation and photoionization reactions found by the sensitivity
analysis are involved in either the formation of key species (C$^+$ and CN), their destruction
(CO, CN, CS, NH$_3$, and HCN), or the destruction of parental molecules (OH, CH, and CH$_2$).
Their rates in the RATE\,06 database are in general inaccurate by a factor of 2 with some exceptions.
The photodissociation of CO proceeds indirectly, through predissociation,
and depends on the amount of molecular hydrogen and CO molecules in the line of sight to the source of UV
radiation \citep{van_Dishoeck87}, and thus has a rate uncertainty of a factor of 10 (see Table~\ref{imp_reac}).
In contrast, the rates of direct photodissociation of NH$_3$, OH, and HCN are inaccurate by the factor of 1.5
only \citep{RJLD91}. Despite such small error bars, the latter three reactions
were identified as problematic reactions for disk chemistry.


\subsubsection{Ion-neutral and neutral-neutral reactions}
\label{INNN_imp}
Ion-neutral (IN) and neutral-neutral (NN)
reactions form the largest fraction of the problematic reactions (32
out of 56) for disk chemistry, but their individual importance is
not as significant as for CRI and RA reactions
(Table~\ref{imp_reac}).

Ion-neutral reactions are usually exothermic and rapid, and
lead to the formation of new species by bond rearrangement. Their
rate coefficients can often be obtained by simple Langevin theory,
which relates the isotropic polarizability of the species and its
reduced mass \citep{Clary_88}. For molecules that possess a large
dipole moment, the rate values can be significantly enhanced at low
temperature by the long-distance Coulomb attraction between the
positive ion and negatively charged side of the molecule
\citep{ASC85,Clary85}.
Moreover, many neutral-neutral reactions with small barriers or without a barrier can also be fast
under interstellar conditions. The rates of most radical-radical reactions and even some radical-stable neutral
reactions are controlled by long-range attractive forces and thus do not get smaller at low temperatures
\citep[see, e.g.,][]{Smith_88,SQTea_93,SSCea94,CSTea97}.
The typical rate coefficient for these reactions is about an order
of magnitude lower than for ion-neutral processes. The rates
of some of the fast neutral-neutral reactions at low temperatures
have been measured in the laboratory and predicted theoretically
\citep[e.g.,][]{Clary_ea94,SQDea94,ChLPSea01,OSU03}.

In RATE\,06, most of the IN and NN rate coefficients are uncertain
by a factor of 2 and less, up to the accuracy of $25\%$
(Table~\ref{imp_reac}). However, a major ambiguity that resides in
the IN rate constants is the treatment of their temperature
dependence. In two versions of the RATE\,06 database the rate value
for the reaction involving a polar molecule and a molecular ion can
have a negative dependence on temperature ($\beta = -0.5$; adopted
in this work) or no temperature dependence ($\beta=0$), while its
rate coefficient fits the value at room temperature. In contrast, in
the OSU network the corresponding IN rates based on low
temperature theoretical estimations to be most accurate at $\sim 10$~K
even though the listed reaction rate coefficients are scaled with respect
to 300 K with $\beta \sim -0.5$. Two problematic
reactions involving a polar molecule and an ion with significantly
different rates in RATE\,06 and osu.2007 are those between C$^+$ and
NH, forming CN$^+$ and H ($\alpha_{\rm osu.2007} = 4.6 \times
10^{-9}$~cm$^3$\,s$^{-1}$ vs. $\alpha_{\rm RATE\,06} = 7.8 \times
10^{-10}$~cm$^3$\,s$^{-1}$) and between S$^+$ and CH, producing
CS$^+$ and H ($\alpha_{\rm osu.2007} = 4.4 \times
10^{-9}$~cm$^3$\,s$^{-1}$ vs $\alpha_{\rm RATE\,06} = 6.2 \times
10^{-10}$~cm$^3$\,s$^{-1}$).

Other IN reactions with different rates in the RATE\,06 and osu.2007
databases are reactions involving an atom and a molecular ion.
The break up of molecular hydrogen upon collision with helium ion
involves a small barrier of 35~K and $\alpha = 3.7\times
10^{-14}$~cm$^3$\,s$^{-1}$ in RATE\,06
\citep[the relevant temperature range is between 10K and 300K;][]{rate06}, while in
osu.2007 the same rate coefficient is about 5 times smaller and has no barrier
\citep{OSU03}. However, this results in difference of the osu.2007 and RATE\,06 rate values
of an order of magnitude at most.

Similarly, the OSU reaction between molecular
hydrogen and ionized ammonia produces hydrogen atom and protonated
ammonia with a rate coefficient of $1.5 \times
10^{-14}$~cm$^3$\,s$^{-1}$ and a steep temperature dependence
($\beta=-1.5$). In RATE\,06 this reaction has a non-Langevin rate with $\beta = 0$.
At temperatures below 20~K its rate coefficient is twice as large, $\alpha = 3.36\times 10^{-14}$~cm$^3$\,s$^{-1}$,
and a negative activation barrier of -35.7~K exists \citep{AS_84}. At temperatures between 20 and 300~K
this reaction has no barrier and $\alpha = 2\times 10^{-13}$~cm$^3$\,s$^{-1}$. For temperatures above 300~K the
rate coefficient is $1.7 \times 10^{-11}$~cm$^3$\,s$^{-1}$ and a barrier is about 1\,000~K
\citep{Fehsenfeld_ea75}. Consequently, the osu.2007 and RATE\,06 rate values can differ by up to one order
of magnitude.

Finally, the IN
reaction between H$_3^+$ and O produces only OH$^+$ and H$_2$ in
OSU, but in RATE\,06 a second, less probable alternative channel
that leads to the production of ionized water and atomic hydrogen is
given \citep{MMcE_00}.

Note that the 4 IN reactions with the smallest rate uncertainty of
$25\%$ are still problematic for disk
chemistry. Such high sensitivity of the disk modeling results to the rates of these
reactions is caused by their importance for the
evolution of CO, HCO$^+$, and HCN.
The most important one is the IN reaction between helium ions and CO, which leads to the hypersensitivity of
the final abundances of many carbon-bearing species in the inner part of the disk
intermediate layer (see Fig.~\ref{hists}).
The other reactions include the primal route to the formation of HCO$^+$ and destruction of CO:
H$_3^+$ + CO $\rightarrow$ HCO$^+$ + H$_2$ \citep[$\alpha = 1.7 \times 10^{-9}$~cm$^3$\,s$^{-1}$;][]{Kim_ea75}
as well as the destruction of molecular nitrogen by ionized helium
\citep[$\alpha = 6.4 \times 10^{-10}$~cm$^3$\,s$^{-1}$;][]{rate06} and
destruction of HCN by ionized carbon
\citep[$\alpha = 3.1 \times 10^{-9}$~cm$^3$\,s$^{-1}$, $\beta=-0.5$;][]{Clary_ea85}.

All of the 15 problematic neutral-neutral reactions for disk chemistry are reactions involving atoms
(O, N, C, H, S) and either light hydrocarbons (C$_{\rm n}$H$_{\rm m}$, n=1,2, m=1--3) or CN, HCO, and H$_2$CO.
In contrast to the IN reactions, the NN reactions are mainly important for the chemical evolution of the key
species in warm disk regions, in particular reaction with atomic oxygen \citep{TG07}.
Though their rates are rather accurate, with uncertainties that are typically not higher than $50\%$,
some of the NN rates are still different in RATE\,06 and osu.2007 databases.

The NN reaction between N and HCO produces only OCN and H in osu.2007, with a rate coefficient of
$10^{-10}$~cm$^3$\,s$^{-1}$ \citep{OSU03}. In RATE\,06 this reaction has in addition 2 other channels
that form either HCN and O ($\alpha = 1.7 \times 10^{-10}$~cm$^3$\,s$^{-1}$)
or CO and NH \citep[$\alpha = 5.7 \times 10^{-12}$~cm$^3$\,s$^{-1}$, $\beta=0.5$, and $\gamma =1\,000$~K;][]{rate06}.

Another reaction in the list is H + CH, which leads to C and H$_2$.
In osu.2007, the rate coefficient has a weak dependence on temperature and no barrier
\citep[$\alpha = 2.7 \times 10^{-11}$~cm$^3$\,s$^{-1}$, $\beta=0.38$;][]{umist99},
while in RATE\,06 this rate has $\beta=0$ and an $80$~K barrier
\citep[$\alpha = 1.3 \times 10^{-10}$~cm$^3$\,s$^{-1}$;][]{rate06}.

Finally, the neutral-neutral reaction between H and CH$_2$,
which forms CH and H$_2$, has the osu.2007 rate coefficient of
$2.7 \times 10^{-10}$~cm$^3$\,s$^{-1}$ taken from the UMIST\,95 database,
but this value is only $6.6 \times 10^{-11}$~cm$^3$\,s$^{-1}$ in RATE\,06.

We conclude that the reaction rates of many important
ion-neutral and neutral-neutral reactions are not known accurately
enough, as a major controversy in their temperature dependence still
persists. This is particularly true for the reactions involving an
ion and a polar molecule, whose rates in RATE\,06 can differ by
a factor of $\sim 5$ at 10~K for the dipole/non-dipole versions of this ratefile.
The ion-polar rates in osu.2007 can be larger than those in the dipole
version of RATE06 by another factor of 4-5, especially for linear neutral
reactants,  in which allowance is made for sub-thermal rotational populations
via the so-called ``locked-dipole'' approach.

\subsubsection{Dissociative recombination}
\label{DR_imp}
Molecular ions are efficiently converted into other, less complex molecules via dissociative
recombination (DR) with electrons or charged grains.
Similarly to IN reactions, these processes are especially fast at low temperatures
with typical rate coefficients of about $10^{-6}$~cm$^{-3}$\,s$^{-1}$ at $10$~K due to the long-range Coulomb
attraction. The DR rates can be measured rather accurately with afterglow and storage ring techniques
\citep{Florescu_Mitchell06}. However, the sensitivity of their branching ratios with respect to the
temperature is mostly unexplored and may not be described by a simple power law at low $T$
\citep{Petrignani_ea05}.

A strong influence of the DR rates and branching ratios on the results of chemical models of molecular clouds
has been found \citep{Millar_ea88,Semaniak_ea01,Geppert_ea05a}. Our sensitivity method picked up two
DR reactions that introduce significant uncertainties in computed abundances in protoplanetary disks.
These include dissociative
recombination of protonated ammonia into ammonia and hydrogen atoms (50th place out of 56), and a main
destruction channel for HCO$^+$ (19th place, see Table~\ref{imp_reac}).
Note that their rates are supposed to be known within an uncertainty of $25\%$.

The DR of HCO$^+$ in RATE\,06 has one dissociative channel to CO and H with the total
rate coefficient of $2.4 \times 10^{-7}$~cm$^3$\,s$^{-1}$ and $\beta=-0.69$ \citep{Mitchell_90}.
In the osu.2007 database 2 additional channels are listed, which lead to the formation of either OH and C or CH and O but
with a much smaller probability of $\sim 4\%$ each. The DR of NH$_4^+$ in both RATE\,06 and osu.2007
involves 3 branching channels into NH$_2$ and either two hydrogen atoms ($21\%$) or one molecular hydrogen
($10\%$), or ammonia and atomic hydrogen ($69\%$) with the total rate of $1.5 \times 10^{-6}$~cm$^3$\,s$^{-1}$
and $\beta=-0.5$ \citep{Vikor_ea99}.

Still, more effort needs to be invested into the (re-)investigation of the DR branching ratios and their
temperature-dependence, especially at low $\sim10$--20~K temperatures.

\section{Summary and conclusions}
\label{concl} The influence of gas-phase reaction rate uncertainties
on the results of disk chemical modeling has been studied by a Monte
Carlo method. The rate coefficients in the RATE\,06 network were
varied 8\,000 times within their uncertainty limits using a
log-normal distribution. Sets of abundances, column densities, and
their error bars were computed for a number of key species in
protoplanetary disks. We found that typical uncertainties of the
molecular column densities do not exceed a factor of 3--4 even for
the largest of the key molecules, which is comparable with
observational uncertainties. The column densities of CO, C$^+$,
H$_3^+$, H$_2$O, NH$_3$, N$_2$H$^+$, and HCNH$^+$ have particularly
small error bars. A straightforward correlation analysis between
molecular abundances and reaction rates was performed for the entire
evolutionary time, and the most problematic reactions involving CO,
C$^+$, CS, HCO$^+$, H$_2$CO, CN, HCN, and NH$_3$ were identified. We
showed that the rate coefficients of about a hundred chemical
reactions constituting only a few percent of the entire RATE\,06
database need to be determined more accurately in order to
significantly decrease uncertainties in the modeled abundances and
column densities of the key observable molecules. We argue that it
is worthwhile to (re-)investigate, either experimentally or
theoretically, the rate coefficients of basic radiative association
and cosmic ray ionization reactions as well as the
temperature-dependence of key ion-neutral and neutral-neutral
reactions, and the branching ratios and products of dissociative
recombination processes. The rate uncertainties of cosmic ray
ionization and photoreactions are partly due to uncertain physical
parameters such as the CRP flux and UV penetration, and thus can be
decreased with better physical models and observational data.

\begin{acknowledgements}
A.V. appreciates financial support through a grant from the Dynasty
Foundation and IMPRS fellowship. E. H. acknowledges the support of the National Science Foundation for
his research program in astrochemistry. A. S. acknowledges the support of the Russian Foundation for Basic Research
(grant 07-02-00628-a). Authors are thankful to the anonymous referee for valuable comments and suggestions.
This research has made use of NASA's Astrophysics Data System.
\end{acknowledgements}


\begin{thebibliography}{115}
\expandafter\ifx\csname natexlab\endcsname\relax\def\natexlab#1{#1}\fi

\bibitem[{{Adams} \& {Smith}(1984)}]{AS_84}
{Adams}, N.~G. \& {Smith}, D. 1984, Int. J. Mass Spectrom. Ion Proc., 61, 133

\bibitem[{{Adams} {et~al.}(1985){Adams}, {Smith}, \& {Clary}}]{ASC85}
{Adams}, N.~G., {Smith}, D., \& {Clary}, D.~C. 1985, \apjl, 296, L31

\bibitem[{{Aikawa} \& {Herbst}(1999)}]{Aikawa_Herbst99}
{Aikawa}, Y. \& {Herbst}, E. 1999, \aap, 351, 233

\bibitem[{{Aikawa} {et~al.}(2003){Aikawa}, {Momose}, {Thi}, {van Zadelhoff},
  {Qi}, {Blake}, \& {van Dishoeck}}]{Aikawa_ea03}
{Aikawa}, Y., {Momose}, M., {Thi}, W.-F., {van Zadelhoff}, G.-J., {Qi}, C.,
  {Blake}, G.~A., \& {van Dishoeck}, E.~F. 2003, \pasj, 55, 11

\bibitem[{{Aikawa} \& {Nomura}(2006)}]{Aikawa_ea06}
{Aikawa}, Y. \& {Nomura}, H. 2006, \apj, 642, 1152

\bibitem[{{Aikawa} {et~al.}(2002){Aikawa}, {van Zadelhoff}, {van Dishoeck}, \&
  {Herbst}}]{Aikawa_ea02}
{Aikawa}, Y., {van Zadelhoff}, G.~J., {van Dishoeck}, E.~F., \& {Herbst}, E.
  2002, \aap, 386, 622

\bibitem[{{Bates}(1951)}]{Bates_51}
{Bates}, D.~R. 1951, \mnras, 111, 303

\bibitem[{{Bates} \& {Herbst}(1988)}]{Bates_Herbst88a}
{Bates}, D.~R. \& {Herbst}, E. 1988, in ASSL Vol. 146: Rate Coefficients in
  Astrochemistry, ed. T.~J. {Millar} \& D.~A. {Williams} (Kluwer Academic
  Publishers, Dordrecht), 17--37

\bibitem[{{Bergin} {et~al.}(2003){Bergin}, {Calvet}, {D'Alessio}, \&
  {Herczeg}}]{Bea03}
{Bergin}, E., {Calvet}, N., {D'Alessio}, P., \& {Herczeg}, G.~J. 2003, \apjl,
  591, L159

\bibitem[{{Bisschop} {et~al.}(2006){Bisschop}, {Fraser}, {{\"O}berg}, {van
  Dishoeck}, \& {Schlemmer}}]{Bisschop_etal2006}
{Bisschop}, S.~E., {Fraser}, H.~J., {{\"O}berg}, K.~I., {van Dishoeck}, E.~F.,
  \& {Schlemmer}, S. 2006, \aap, 449, 1297

\bibitem[{{Boger} \& {Sternberg}(2006)}]{Boger_Sternberg06}
{Boger}, G.~I. \& {Sternberg}, A. 2006, \apj, 645, 314

\bibitem[{{Brownsword} {et~al.}(1997){Brownsword}, {Sims}, {Smith}, {Stewart},
  {Canosa}, \& {Rowe}}]{BS97}
{Brownsword}, R.~A., {Sims}, I.~R., {Smith}, I.~W.~M., {Stewart}, D.~W.~A.,
  {Canosa}, A., \& {Rowe}, B.~R. 1997, \apj, 485, 195

\bibitem[{{Canosa} {et~al.}(1997){Canosa}, {Sims}, {Travers}, {Smith}, \&
  {Rowe}}]{CSTea97}
{Canosa}, A., {Sims}, I.~R., {Travers}, D., {Smith}, I.~W.~M., \& {Rowe}, B.~R.
  1997, \aap, 323, 644

\bibitem[{{Caselli} {et~al.}(1998){Caselli}, {Walmsley}, {Terzieva}, \&
  {Herbst}}]{Caselli_ea98}
{Caselli}, P., {Walmsley}, C.~M., {Terzieva}, R., \& {Herbst}, E. 1998, \apj,
  499, 234

\bibitem[{{Chastaing} {et~al.}(2001){Chastaing}, {Le Picard}, {Sims}, \&
  {Smith}}]{ChLPSea01}
{Chastaing}, D., {Le Picard}, S.~D., {Sims}, I.~R., \& {Smith}, I.~W.~M. 2001,
  \aap, 365, 241

\bibitem[{{Clary}(1985)}]{Clary85}
{Clary}, D.~C. 1985, Molec. Phys., 54, 605

\bibitem[{{Clary}(1988)}]{Clary_88}
---. 1988, in ASSL Vol. 146: Rate Coefficients in Astrochemistry, ed. T.~J.
  {Millar} \& D.~A. {Williams} (Kluwer Academic Publishers, Dordrecht), 1--24

\bibitem[{{Clary} {et~al.}(1994){Clary}, {Haider}, {Husain}, \&
  {Kabir}}]{Clary_ea94}
{Clary}, D.~C., {Haider}, N., {Husain}, D., \& {Kabir}, M. 1994, \apj, 422, 416

\bibitem[{{Clary} {et~al.}(1985){Clary}, {Smith}, \& {Adams}}]{Clary_ea85}
{Clary}, D.~C., {Smith}, D., \& {Adams}, N.~G. 1985, Chemical Physics Letters,
  119, 320

\bibitem[{{D'Alessio} {et~al.}(1999){D'Alessio}, {Calvet}, {Hartmann},
  {Lizano}, \& {Cant{\'o}}}]{DAlessio_ea99}
{D'Alessio}, P., {Calvet}, N., {Hartmann}, L., {Lizano}, S., \& {Cant{\'o}}, J.
  1999, \apj, 527, 893

\bibitem[{{Dalgarno} \& {McCray}(1972)}]{DalgarnoMacCray72}
{Dalgarno}, A. \& {McCray}, R.~A. 1972, \araa, 10, 375

\bibitem[{{Dalgarno} \& {Stephens}(1970)}]{DS70}
{Dalgarno}, A. \& {Stephens}, T.~L. 1970, \apjl, 160, L107

\bibitem[{{Dartois} {et~al.}(2003){Dartois}, {Dutrey}, \&
  {Guilloteau}}]{Dartois_ea03}
{Dartois}, E., {Dutrey}, A., \& {Guilloteau}, S. 2003, \aap, 399, 773

\bibitem[{{Dobrijevic} {et~al.}(2003){Dobrijevic}, {Ollivier}, {Billebaud},
  {Brillet}, \& {Parisot}}]{Dobrijevic_ea03}
{Dobrijevic}, M., {Ollivier}, J.~L., {Billebaud}, F., {Brillet}, J., \&
  {Parisot}, J.~P. 2003, \aap, 398, 335

\bibitem[{{Dobrijevic} \& {Parisot}(1998)}]{Dobrijevic_Parisot98}
{Dobrijevic}, M. \& {Parisot}, J.~P. 1998, \planss, 46, 491

\bibitem[{{Dolginov} \& {Stepinski}(1994)}]{DolginovStepinski94}
{Dolginov}, A.~Z. \& {Stepinski}, T.~F. 1994, \apj, 427, 377

\bibitem[{{Draine}(1978)}]{Draine_78}
{Draine}, B.~T. 1978, \apjs, 36, 595

\bibitem[{{Dutrey} {et~al.}(1997){Dutrey}, {Guilloteau}, \&
  {Guelin}}]{Dutrey_ea97}
{Dutrey}, A., {Guilloteau}, S., \& {Guelin}, M. 1997, \aap, 317, L55

\bibitem[{{Dutrey} {et~al.}(2007){Dutrey}, {Henning}, {Guilloteau}, {Semenov},
  {Pi{\'e}tu}, {Schreyer}, {Bacmann}, {Launhardt}, {Pety}, \&
  {Gueth}}]{Dutrey_ea07}
{Dutrey}, A., {Henning}, T., {Guilloteau}, S., {Semenov}, D., {Pi{\'e}tu}, V.,
  {Schreyer}, K., {Bacmann}, A., {Launhardt}, R., {Pety}, J., \& {Gueth}, F.
  2007, \aap, 464, 615

\bibitem[{{Fehsenfeld} {et~al.}(1975){Fehsenfeld}, {Lindinger}, {Schmeltekopf},
  {Albritton}, \& {Ferguson}}]{Fehsenfeld_ea75}
{Fehsenfeld}, F.~C., {Lindinger}, W., {Schmeltekopf}, A.~L., {Albritton},
  D.~L., \& {Ferguson}, E.~E. 1975, \jcp, 62, 2001

\bibitem[{{Finocchi} \& {Gail}(1997)}]{Finocchi_Gail97}
{Finocchi}, F. \& {Gail}, H.-P. 1997, \aap, 327, 825

\bibitem[{{Florescu-Mitchell} \& {Mitchell}(2006)}]{Florescu_Mitchell06}
{Florescu-Mitchell}, A.~I. \& {Mitchell}, J.~B.~A. 2006, \physrep, 430, 277

\bibitem[{{Gammie}(1996)}]{Gammie96}
{Gammie}, C.~F. 1996, \apj, 457, 355

\bibitem[{{Garrod} \& {Herbst}(2006)}]{GH_06}
{Garrod}, R.~T. \& {Herbst}, E. 2006, \aap, 457, 927

\bibitem[{{Geppert} {et~al.}(2005a){Geppert}, {Thomas}, {Ehlerding},
  {Hellberg}, {{\"O}sterdahl}, {Hamberg}, {Semaniak}, {Zhaunerchyk},
  {Kaminska}, {K{\"a}llberg}, {Paal}, \& {Larsson}}]{Geppert_ea05a}
{Geppert}, W.~D., {Thomas}, R.~D., {Ehlerding}, A., {Hellberg}, F.,
  {{\"O}sterdahl}, F., {Hamberg}, M., {Semaniak}, J., {Zhaunerchyk}, V.,
  {Kaminska}, M., {K{\"a}llberg}, A., {Paal}, A., \& {Larsson}, M. 2005a,
  Journal of Physics Conference Series, 4, 26

\bibitem[{{Gerlich}(1995)}]{G95}
{Gerlich}, D. 1995, Physica Scripta, 1995, 256

\bibitem[{{Glassgold} {et~al.}(2005){Glassgold}, {Feigelson}, {Montmerle}, \&
  {Wolk}}]{Glassgold_ea05}
{Glassgold}, A.~E., {Feigelson}, E.~D., {Montmerle}, T., \& {Wolk}, S. 2005, in
  ASP Conf. Ser. 341: Chondrites and the Protoplanetary Disk, ed. A.~N. {Krot},
  E.~R.~D. {Scott}, \& B.~{Reipurth}, 165--178

\bibitem[{{Glassgold} \& {Langer}(1973)}]{Glassgold_Langer_73}
{Glassgold}, A.~E. \& {Langer}, W.~D. 1973, \apj, 186, 859

\bibitem[{{Glassgold} {et~al.}(1997{\natexlab{a}}){Glassgold}, {Najita}, \&
  {Igea}}]{Glassgold_ea97a}
{Glassgold}, A.~E., {Najita}, J., \& {Igea}, J. 1997{\natexlab{a}}, \apj, 480,
  344

\bibitem[{{Glassgold} {et~al.}(1997{\natexlab{b}}){Glassgold}, {Najita}, \&
  {Igea}}]{Glassgold_ea97b}
---. 1997{\natexlab{b}}, \apj, 485, 920

\bibitem[{{Hasegawa} \& {Herbst}(1993)}]{Hasegawa_Herbst93}
{Hasegawa}, T.~I. \& {Herbst}, E. 1993, \mnras, 263, 589

\bibitem[{{Herbst}(1980)}]{Herbst_80}
{Herbst}, E. 1980, \apj, 241, 197

\bibitem[{{Herbst}(1985)}]{Herbst_85}
---. 1985, \apj, 291, 226

\bibitem[{{Herbst} \& {Klemperer}(1973)}]{Herbst_Klemperer73}
{Herbst}, E. \& {Klemperer}, W. 1973, \apj, 185, 505

\bibitem[{{Hollenbach} \& {McKee}(1979)}]{Hollenbach_McKee79}
{Hollenbach}, D. \& {McKee}, C.~F. 1979, \apjs, 41, 555

\bibitem[{{Hollis} {et~al.}(2006){Hollis}, {Lovas}, {Remijan}, {Jewell},
  {Ilyushin}, \& {Kleiner}}]{Hollis_ea06}
{Hollis}, J.~M., {Lovas}, F.~J., {Remijan}, A.~J., {Jewell}, P.~R., {Ilyushin},
  V.~V., \& {Kleiner}, I. 2006, \apjl, 643, L25

\bibitem[{{Ilgner} {et~al.}(2004){Ilgner}, {Henning}, {Markwick}, \&
  {Millar}}]{Ilgner_ea04}
{Ilgner}, M., {Henning}, T., {Markwick}, A.~J., \& {Millar}, T.~J. 2004, \aap,
  415, 643

\bibitem[{{Ilgner} \& {Nelson}(2006)}]{Ilgner_ea06a}
{Ilgner}, M. \& {Nelson}, R.~P. 2006, \aap, 445, 205

\bibitem[{{Ishii} {et~al.}(2006){Ishii}, {Tajima}, {Taketsugu}, \&
  {Yamashita}}]{Ishii_ea06}
{Ishii}, K., {Tajima}, A., {Taketsugu}, T., \& {Yamashita}, K. 2006, \apj, 636,
  927

\bibitem[{{Izzard} {et~al.}(2007){Izzard}, {Lugaro}, {Karakas}, {Iliadis}, \&
  {van Raai}}]{Izzard_ea07}
{Izzard}, R.~G., {Lugaro}, M., {Karakas}, A.~I., {Iliadis}, C., \& {van Raai},
  M. 2007, \aap, 466, 641

\bibitem[{{Jonkheid} {et~al.}(2006){Jonkheid}, {Kamp}, {Augereau}, \& {van
  Dishoeck}}]{Jonkheid_ea07}
{Jonkheid}, B., {Kamp}, I., {Augereau}, J.-C., \& {van Dishoeck}, E.~F. 2006,
  \aap, 453, 163

\bibitem[{{Kastner} {et~al.}(1997){Kastner}, {Zuckerman}, {Weintraub}, \&
  {Forveille}}]{Kastner_etal1997}
{Kastner}, J.~H., {Zuckerman}, B., {Weintraub}, D.~A., \& {Forveille}, T. 1997,
  Science, 277, 67

\bibitem[{{Kim} {et~al.}(1975){Kim}, {Theard}, \& {Huntress}}]{Kim_ea75}
{Kim}, J.~K., {Theard}, L.~P., \& {Huntress}, Jr., W.~T. 1975, Chemical Physics
  Letters, 32, 610

\bibitem[{{Le Teuff} {et~al.}(2000){Le Teuff}, {Millar}, \&
  {Markwick}}]{umist99}
{Le Teuff}, Y.~H., {Millar}, T.~J., \& {Markwick}, A.~J. 2000, \aaps, 146, 157

\bibitem[{{Lee} {et~al.}(1996){Lee}, {Herbst}, {Pineau des Forets}, {Roueff},
  \& {Le Bourlot}}]{Lee_ea96}
{Lee}, H.-H., {Herbst}, E., {Pineau des Forets}, G., {Roueff}, E., \& {Le
  Bourlot}, J. 1996, \aap, 311, 690

\bibitem[{{Lee} {et~al.}(1998){Lee}, {Roueff}, {Pineau des Forets},
  {Shalabiea}, {Terzieva}, \& {Herbst}}]{Lee_ea98}
{Lee}, H.-H., {Roueff}, E., {Pineau des Forets}, G., {Shalabiea}, O.~M.,
  {Terzieva}, R., \& {Herbst}, E. 1998, \aap, 334, 1047

\bibitem[{{Leger} {et~al.}(1985){Leger}, {Jura}, \& {Omont}}]{LJO85}
{Leger}, A., {Jura}, M., \& {Omont}, A. 1985, \aap, 144, 147

\bibitem[{{Lepp} \& {Dalgarno}(1996)}]{LeppDalgarno97}
{Lepp}, S. \& {Dalgarno}, A. 1996, \aap, 306, L21

\bibitem[{{Maloney} {et~al.}(1996){Maloney}, {Hollenbach}, \&
  {Tielens}}]{Maloneyea96}
{Maloney}, P.~R., {Hollenbach}, D.~J., \& {Tielens}, A.~G.~G.~M. 1996, \apj,
  466, 561

\bibitem[{{Markwick} {et~al.}(2002){Markwick}, {Ilgner}, {Millar}, \&
  {Henning}}]{Markwick_ea02}
{Markwick}, A.~J., {Ilgner}, M., {Millar}, T.~J., \& {Henning}, T. 2002, \aap,
  385, 632

\bibitem[{{Millar} {et~al.}(1988){Millar}, {Defrees}, {McLean}, \&
  {Herbst}}]{Millar_ea88}
{Millar}, T.~J., {Defrees}, D.~J., {McLean}, A.~D., \& {Herbst}, E. 1988, \aap,
  194, 250

\bibitem[{{Millar} {et~al.}(1997){Millar}, {Farquhar}, \& {Willacy}}]{umist95}
{Millar}, T.~J., {Farquhar}, P.~R.~A., \& {Willacy}, K. 1997, \aaps, 121, 139

\bibitem[{{Milligan} \& {McEwan}(2000)}]{MMcE_00}
{Milligan}, D.~B. \& {McEwan}, M.~J. 2000, Chemical Physics Letters, 319, 482

\bibitem[{{Mitchell}(1990)}]{Mitchell_90}
{Mitchell}, J.~B.~A. 1990, \physrep, 186, 215

\bibitem[{{Nahar} \& {Pradhan}(1997)}]{Nahar_Pradhan97}
{Nahar}, S.~N. \& {Pradhan}, A.~K. 1997, \apjs, 111, 339

\bibitem[{{Neufeld} {et~al.}(1994){Neufeld}, {Maloney}, \&
  {Conger}}]{Neufeldea94}
{Neufeld}, D.~A., {Maloney}, P.~R., \& {Conger}, S. 1994, \apjl, 436, L127

\bibitem[{{{\"O}jekull} {et~al.}(2006){{\"O}jekull}, {Andersson}, {Nagard},
  {Pettersson}, {Neau}, {Ros{\'e}n}, {Thomas}, {Larsson}, {Semaniak},
  {{\"O}sterdahl}, {Danared}, {K{\"a}llberg}, \& {Ugglas}}]{Ojekull_ea06}
{{\"O}jekull}, J., {Andersson}, P.~U., {Nagard}, M.~B., {Pettersson}, J.~B.~C.,
  {Neau}, A., {Ros{\'e}n}, S., {Thomas}, R.~D., {Larsson}, M., {Semaniak}, J.,
  {{\"O}sterdahl}, F., {Danared}, H., {K{\"a}llberg}, A., \& {Ugglas}, M.~A.
  2006, \jcp, 125, 4306

\bibitem[{{Petrignani} {et~al.}(2005){Petrignani}, {van der Zande}, {Cosby},
  {Hellberg}, {Thomas}, \& {Larsson}}]{Petrignani_ea05}
{Petrignani}, A., {van der Zande}, W.~J., {Cosby}, P.~C., {Hellberg}, F.,
  {Thomas}, R.~D., \& {Larsson}, M. 2005, \jcp, 122, 4302

\bibitem[{{Pety} {et~al.}(2006){Pety}, {Gueth}, {Guilloteau}, \&
  {Dutrey}}]{Pety_ea06}
{Pety}, J., {Gueth}, F., {Guilloteau}, S., \& {Dutrey}, A. 2006, \aap, 458, 841

\bibitem[{{Pi{\'e}tu} {et~al.}(2007){Pi{\'e}tu}, {Dutrey}, \&
  {Guilloteau}}]{Pietu_ea07}
{Pi{\'e}tu}, V., {Dutrey}, A., \& {Guilloteau}, S. 2007, \aap, 467, 163

\bibitem[{{Pineau des Forets} {et~al.}(1992){Pineau des Forets}, {Roueff}, \&
  {Flower}}]{Pineau_des_Forets_ea92}
{Pineau des Forets}, G., {Roueff}, E., \& {Flower}, D.~R. 1992, \mnras, 258,
  45P

\bibitem[{{Prasad} \& {Huntress}(1980)}]{Prasad_Huntress80}
{Prasad}, S.~S. \& {Huntress}, Jr., W.~T. 1980, \apjs, 43, 1

\bibitem[{{Prasad} \& {Tarafdar}(1983)}]{PrasadTarafdar83}
{Prasad}, S.~S. \& {Tarafdar}, S.~P. 1983, \apj, 267, 603

\bibitem[{{Qi} {et~al.}(2003){Qi}, {Kessler}, {Koerner}, {Sargent}, \&
  {Blake}}]{Qi_etal2003}
{Qi}, C., {Kessler}, J.~E., {Koerner}, D.~W., {Sargent}, A.~I., \& {Blake},
  G.~A. 2003, \apj, 597, 986

\bibitem[{{Qi} {et~al.}(2005){Qi}, {Wilner}, {Blake}, {Bourke}, {Hogerheijde},
  \& {Ho}}]{Qi_ea05}
{Qi}, C., {Wilner}, D.~J., {Blake}, G.~A., {Bourke}, T.~L., {Hogerheijde}, M.,
  \& {Ho}, P.~T.~P. 2005, in IAU Symposium, ed. D.~C. {Lis}, G.~A. {Blake}, \&
  E.~{Herbst}, 188--192

\bibitem[{{Roberge} {et~al.}(1991){Roberge}, {Jones}, {Lepp}, \&
  {Dalgarno}}]{RJLD91}
{Roberge}, W.~G., {Jones}, D., {Lepp}, S., \& {Dalgarno}, A. 1991, \apjs, 77,
  287

\bibitem[{{Semaniak} {et~al.}(2001){Semaniak}, {Minaev}, {Derkatch},
  {Hellberg}, {Neau}, {Ros{\' e}n}, {Thomas}, {Larsson}, {Danared}, {Pa{\'
  a}l}, \& {af Ugglas}}]{Semaniak_ea01}
{Semaniak}, J., {Minaev}, B.~F., {Derkatch}, A.~M., {Hellberg}, F., {Neau}, A.,
  {Ros{\' e}n}, S., {Thomas}, R., {Larsson}, M., {Danared}, H., {Pa{\' a}l},
  A., \& {af Ugglas}, M. 2001, \apjs, 135, 275

\bibitem[{{Semenov} {et~al.}(2003){Semenov}, {Henning}, {Helling}, {Ilgner}, \&
  {Sedlmayr}}]{Semenov_ea03}
{Semenov}, D., {Henning}, T., {Helling}, C., {Ilgner}, M., \& {Sedlmayr}, E.
  2003, \aap, 410, 611

\bibitem[{{Semenov} {et~al.}(2005){Semenov}, {Pavlyuchenkov}, {Schreyer},
  {Henning}, {Dullemond}, \& {Bacmann}}]{Semenov_ea05}
{Semenov}, D., {Pavlyuchenkov}, Y., {Schreyer}, K., {Henning}, T., {Dullemond},
  C., \& {Bacmann}, A. 2005, \apj, 621, 853

\bibitem[{{Semenov} {et~al.}(2004){Semenov}, {Wiebe}, \&
  {Henning}}]{Semenov_ea04}
{Semenov}, D., {Wiebe}, D., \& {Henning}, T. 2004, \aap, 417, 93

\bibitem[{{Shalabiea} \& {Greenberg}(1995)}]{Shalabiea_Greenberg95}
{Shalabiea}, O.~M. \& {Greenberg}, J.~M. 1995, \aap, 296, 779

\bibitem[{{Simon} {et~al.}(2000){Simon}, {Dutrey}, \&
  {Guilloteau}}]{Simon_ea00}
{Simon}, M., {Dutrey}, A., \& {Guilloteau}, S. 2000, \apj, 545, 1034

\bibitem[{{Sims} {et~al.}(1994{\natexlab{a}}){Sims}, {Queffelec}, {Defrance},
  {Rebrion-Rowe}, {Travers}, {Bocherel}, {Rowe}, \& {Smith}}]{SQDea94}
{Sims}, I.~R., {Queffelec}, J.-L., {Defrance}, A., {Rebrion-Rowe}, C.,
  {Travers}, D., {Bocherel}, P., {Rowe}, B.~R., \& {Smith}, I.~W.~M.
  1994{\natexlab{a}}, \jcp, 100, 4229

\bibitem[{{Sims} {et~al.}(1993){Sims}, {Queffelec}, {Travers}, {Rowe},
  {Herbert}, {Karthauser}, \& {Smith}}]{SQTea_93}
{Sims}, I.~R., {Queffelec}, J.-L., {Travers}, D., {Rowe}, B.~R., {Herbert},
  L.~B., {Karthauser}, J., \& {Smith}, I.~W.~M. 1993, Chemical Physics Letters,
  211, 461

\bibitem[{{Sims} {et~al.}(1994{\natexlab{b}}){Sims}, {Smith}, {Clary},
  {Bocherel}, \& {Rowe}}]{SSCea94}
{Sims}, I.~R., {Smith}, I.~W.~M., {Clary}, D.~C., {Bocherel}, P., \& {Rowe},
  B.~R. 1994{\natexlab{b}}, \jcp, 101, 1748

\bibitem[{{Singh} \& {Andreazza}(2000)}]{SA00}
{Singh}, P.~D. \& {Andreazza}, C.~M. 2000, \apj, 537, 261

\bibitem[{{Smith}(1988)}]{Smith_88}
{Smith}, I.~W.~M. 1988, in ASSL Vol. 146: Rate Coefficients in Astrochemistry,
  ed. T.~J. {Millar} \& D.~A. {Williams} (Kluwer Academic Publishers,
  Dordrecht), 106--116

\bibitem[{{Smith}(1989)}]{Smith89}
---. 1989, \apj, 347, 282

\bibitem[{{Smith} {et~al.}(2004){Smith}, {Herbst}, \& {Chang}}]{OSU03}
{Smith}, I.~W.~M., {Herbst}, E., \& {Chang}, Q. 2004, \mnras, 350, 323

\bibitem[{{Snyder} {et~al.}(1974){Snyder}, {Buhl}, {Schwartz}, {Clark},
  {Johnson}, {Lovas}, \& {Giguere}}]{Snyder_ea74}
{Snyder}, L.~E., {Buhl}, D., {Schwartz}, P.~R., {Clark}, F.~O., {Johnson},
  D.~R., {Lovas}, F.~J., \& {Giguere}, P.~T. 1974, \apjl, 191, L79+

\bibitem[{{Solomon} \& {Werner}(1971)}]{Solomon_Werner71}
{Solomon}, P.~M. \& {Werner}, M.~W. 1971, \apj, 165, 41

\bibitem[{{Spitzer}(1978)}]{Spitzer_78}
{Spitzer}, L. 1978, {Physical processes in the interstellar medium} (New York
  Wiley-Interscience, 1978.~333 p.)

\bibitem[{{Spitzer} \& {Tomasko}(1968)}]{Spitzer_Tomasko68}
{Spitzer}, L.~J. \& {Tomasko}, M.~G. 1968, \apj, 152, 971

\bibitem[{{St{\"a}uber} {et~al.}(2005){St{\"a}uber}, {Doty}, {van Dishoeck}, \&
  {Benz}}]{Staeuber_ea05}
{St{\"a}uber}, P., {Doty}, S.~D., {van Dishoeck}, E.~F., \& {Benz}, A.~O. 2005,
  \aap, 440, 949

\bibitem[{{Thi} {et~al.}(2004){Thi}, {van Zadelhoff}, \& {van
  Dishoeck}}]{Thi_etal2004}
{Thi}, W.-F., {van Zadelhoff}, G.-J., \& {van Dishoeck}, E.~F. 2004, \aap, 425,
  955

\bibitem[{{Tscharnuter} \& {Gail}(2007)}]{TG07}
{Tscharnuter}, W.~M. \& {Gail}, H.-P. 2007, \aap, 463, 369

\bibitem[{{Turanyi}(1997)}]{Turanyi_97}
{Turanyi}, T. 1997, Reliability Engineering and System Safety, 57, 41

\bibitem[{{van Dishoeck}(1987)}]{van_Dishoeck87}
{van Dishoeck}, E.~F. 1987, in IAU Symp. 120: Astrochemistry, ed. M.~S.
  {Vardya} \& S.~P. {Tarafdar}, 51--63

\bibitem[{{van Dishoeck}(1988)}]{van_Dishoeck88}
{van Dishoeck}, E.~F. 1988, in ASSL Vol. 146: Rate Coefficients in
  Astrochemistry, ed. T.~J. {Millar} \& D.~A. {Williams}, 49--82

\bibitem[{{van Dishoeck}(1994)}]{vanDishoeck94}
{van Dishoeck}, E.~F. 1994, in ASP Conf. Ser. 58: The First Symposium on the
  Infrared Cirrus and Diffuse Interstellar Clouds, ed. R.~Cutri \& W.~Latter
  (Astr. Soc. Pacific), 319--331

\bibitem[{{van Dishoeck} \& {Blake}(1998)}]{van_Dishoeck_Blake98}
{van Dishoeck}, E.~F. \& {Blake}, G.~A. 1998, \araa, 36, 317

\bibitem[{{van Dishoeck} {et~al.}(2006){van Dishoeck}, {Jonkheid}, \& {van
  Hemert}}]{van_Dishoeck_ea06}
{van Dishoeck}, E.~F., {Jonkheid}, B., \& {van Hemert}, M.~C. 2006, in Faraday
  Discussions 133: Chemical Evolution of the Universe, ed. I.~R. {Sims} \&
  D.~A. {Williams}, 231--243

\bibitem[{{van Zadelhoff} {et~al.}(2003){van Zadelhoff}, {Aikawa},
  {Hogerheijde}, \& {van Dishoeck}}]{van_Zadelhoff_ea03}
{van Zadelhoff}, G.-J., {Aikawa}, Y., {Hogerheijde}, M.~R., \& {van Dishoeck},
  E.~F. 2003, \aap, 397, 789

\bibitem[{{van Zadelhoff} {et~al.}(2001){van Zadelhoff}, {van Dishoeck}, {Thi},
  \& {Blake}}]{Zadelhoff_etal2001}
{van Zadelhoff}, G.-J., {van Dishoeck}, E.~F., {Thi}, W.-F., \& {Blake}, G.~A.
  2001, \aap, 377, 566

\bibitem[{{Vasyunin} {et~al.}(2004){Vasyunin}, {Sobolev}, {Wiebe}, \&
  {Semenov}}]{Vasyunin_ea04}
{Vasyunin}, A.~I., {Sobolev}, A.~M., {Wiebe}, D.~S., \& {Semenov}, D.~A. 2004,
  Astronomy Letters, 30, 566

\bibitem[{{Vikor} {et~al.}(1999){Vikor}, {Al-Khalili}, {Danared}, {Djuric},
  {Dunn}, {Larsson}, {Le Padellec}, {Rosaen}, \& {Af Ugglas}}]{Vikor_ea99}
{Vikor}, L., {Al-Khalili}, A., {Danared}, H., {Djuric}, N., {Dunn}, G.~H.,
  {Larsson}, M., {Le Padellec}, A., {Rosaen}, S., \& {Af Ugglas}, M. 1999,
  \aap, 344, 1027

\bibitem[{{Wakelam} {et~al.}(2006{\natexlab{a}}){Wakelam}, {Herbst}, \&
  {Selsis}}]{Wakelam_ea06}
{Wakelam}, V., {Herbst}, E., \& {Selsis}, F. 2006{\natexlab{a}}, \aap, 451, 551

\bibitem[{{Wakelam} {et~al.}(2006{\natexlab{b}}){Wakelam}, {Herbst}, {Selsis},
  \& {Massacrier}}]{Wakelam_ea06b}
{Wakelam}, V., {Herbst}, E., {Selsis}, F., \& {Massacrier}, G.
  2006{\natexlab{b}}, \aap, 459, 813

\bibitem[{{Wakelam} {et~al.}(2005){Wakelam}, {Selsis}, {Herbst}, \&
  {Caselli}}]{Wakelam_ea05}
{Wakelam}, V., {Selsis}, F., {Herbst}, E., \& {Caselli}, P. 2005, \aap, 444,
  883

\bibitem[{{Wiebe} {et~al.}(2003){Wiebe}, {Semenov}, \& {Henning}}]{Wiebe_ea03}
{Wiebe}, D., {Semenov}, D., \& {Henning}, T. 2003, \aap, 399, 197

\bibitem[{{Willacy} {et~al.}(1998){Willacy}, {Klahr}, {Millar}, \&
  {Henning}}]{Willacy_ea98}
{Willacy}, K., {Klahr}, H.~H., {Millar}, T.~J., \& {Henning}, T. 1998, \aap,
  338, 995

\bibitem[{{Willacy} {et~al.}(2006){Willacy}, {Langer}, {Allen}, \&
  {Bryden}}]{Willacy_ea06}
{Willacy}, K., {Langer}, W., {Allen}, M., \& {Bryden}, G. 2006, \apj, 644, 1202

\bibitem[{{Willacy} \& {Langer}(2000)}]{Willacy_Langer00}
{Willacy}, K. \& {Langer}, W.~D. 2000, \apj, 544, 903

\bibitem[{{Williams}(1972)}]{Williams72}
{Williams}, D.~A. 1972, \aplett, 10, L17

\bibitem[{{Woodall} {et~al.}(2007){Woodall}, {Agundez}, {Markwick-Kemper}, \&
  {Millar}}]{rate06}
{Woodall}, J., {Agundez}, M., {Markwick-Kemper}, A.~J., \& {Millar}, T.~J.
  2007, \aap, 466, 1197

\end{thebibliography}

\clearpage
\begin{figure}
\includegraphics[width=0.28\textwidth,clip=,angle=90]{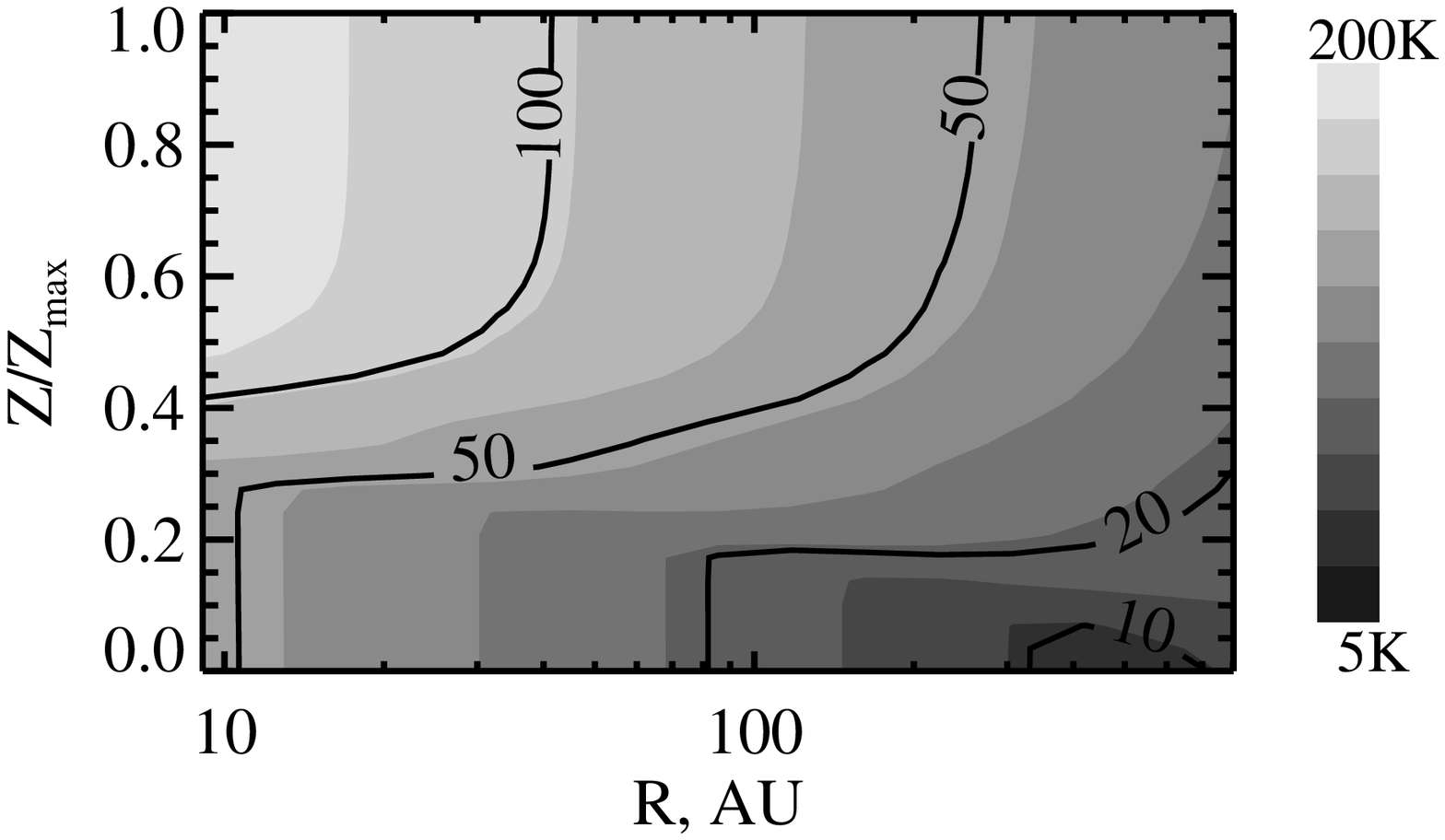}
\includegraphics[width=0.28\textwidth,clip=,angle=90]{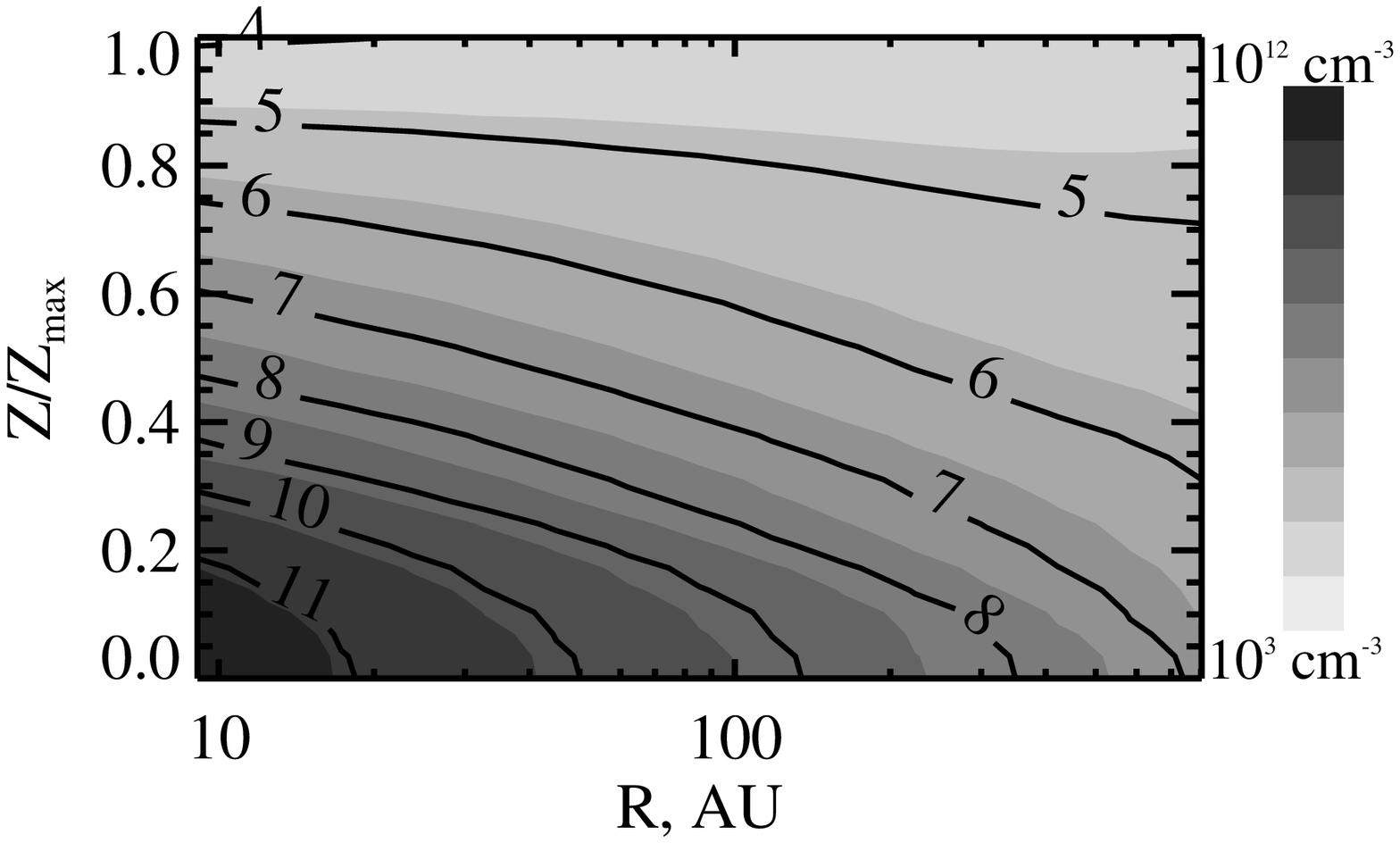}
\caption{Temperature (K) and density (cm$^{-3}$) structure of the adopted disk model.
The total scale height of the disk $Z_{\rm max}$ increases with the radius such that at
$r=100$~AU it has a value of 90~AU, whereas $Z_{\rm max} \approx 1\,300$~AU at $r=800$~AU.}
\label{disk}
\end{figure}

\clearpage
\begin{figure}
\includegraphics[width=0.45\textwidth,clip=]{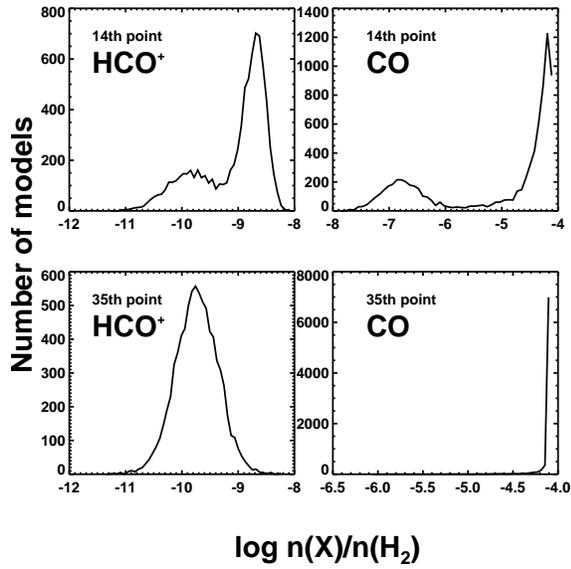}
\caption{Histograms of the abundance distributions show the number of chemical models
in which a given molecular abundance  is reached at 5~Myr. (Top) Non-Gaussian, double-peak
distributions are shown for HCO$^+$ (left) and CO (right) in the inner disk region
($r\approx 97.5$~AU, $z \approx 30$~AU).
(Bottom) Nearly Gaussian distributions are shown for the same species but located
farther out from the central star ($r\approx 382$~AU, $z \approx 205$~AU).
The point numbers in the boxes refer to the disk cells for which these calculations were
performed.}
\label{hists}
\end{figure}

\clearpage
\begin{figure}
\includegraphics[width=0.65\textwidth,clip=,angle=90]{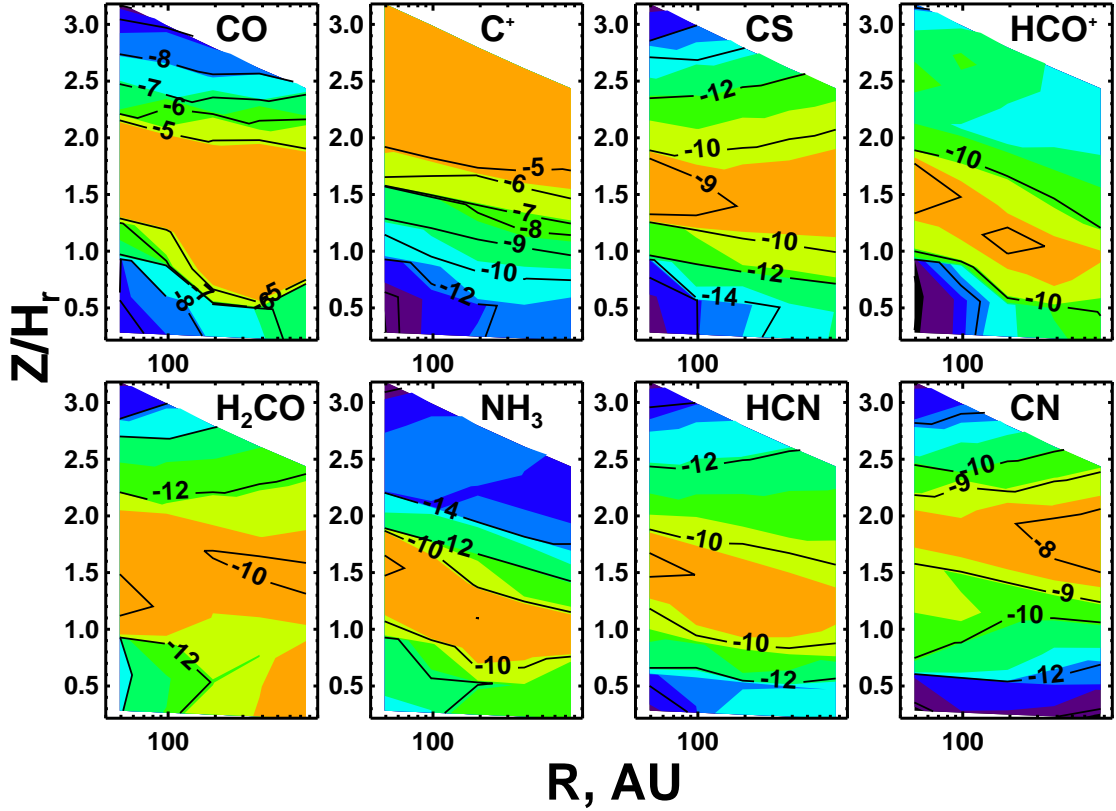}
\caption{Distribution of  mean molecular abundances relative to the total amount of hydrogen nuclei
for several observationally important species in the disk at 5~Myr. The distributions are obtained by
averaging the 8\,000 model results calculated with  randomly varied reaction rates. Results are
shown for CO, C$^+$, CS, HCO$^+$ (top row, from left to right) and H$_2$CO, NH$_3$, HCN, and CN
(bottom row) at $r\ge 50$~AU. The disk vertical axis is given in units of the pressure scale height.}
\label{Abu2D}
\end{figure}

\clearpage
\begin{figure}
\includegraphics[width=0.65\textwidth,clip=,angle=90]{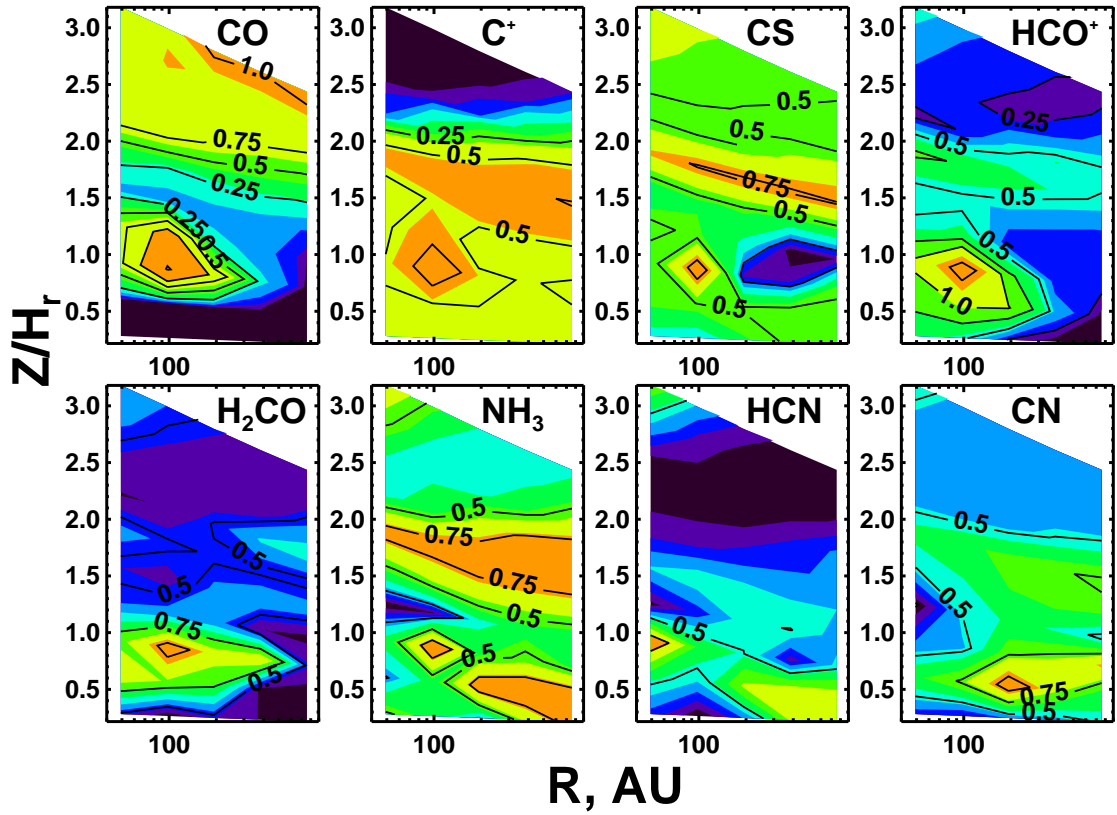}
\caption{The same as in Fig.~\ref{Abu2D} but for abundance uncertainties (logarithmic units).}
\label{UncertDistrib}
\end{figure}

\clearpage
\begin{figure}
\includegraphics[width=0.65\textwidth,clip=,angle=90]{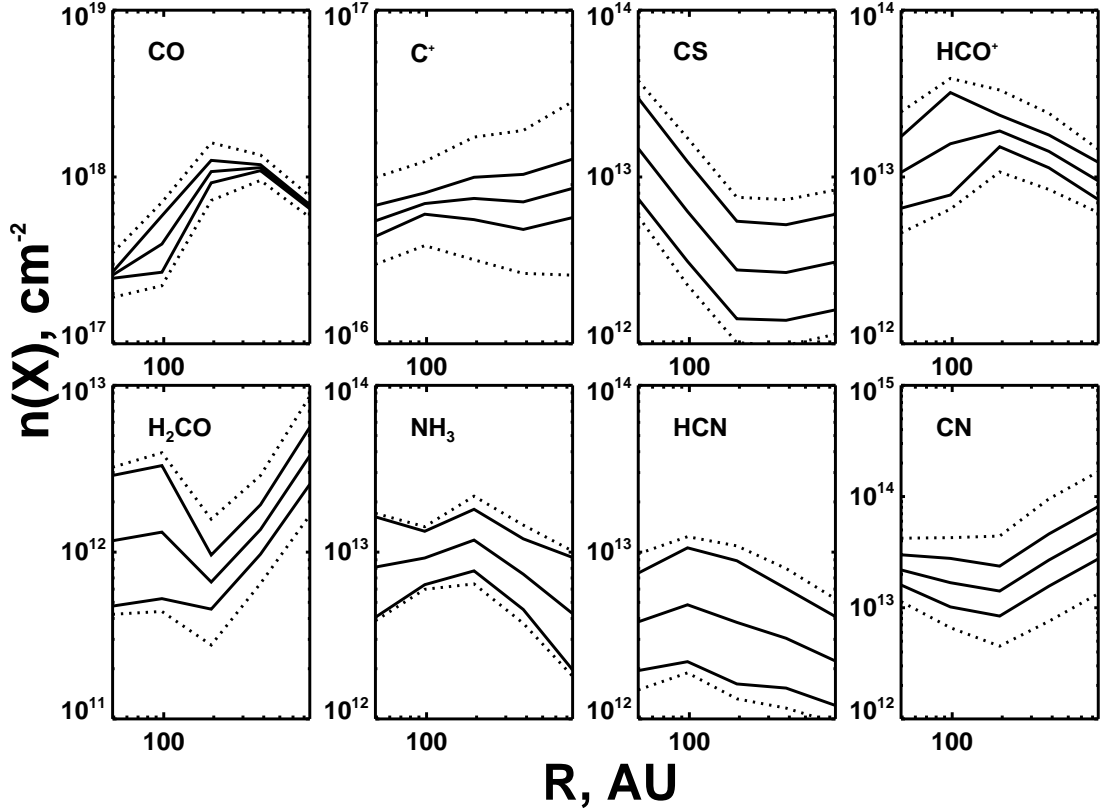}
\caption{The same as in Fig.~\ref{UncertDistrib} but for radial distributions of the column densities.
The central solid line represents the model-averaged column densities.
The corresponding column density scatters computed with the standard model are depicted with dotted lines.
The column density scatters calculated with the chemical model where the rates of the 56 most problematic reactions
are varied by small factors of 1.25--2 only are shown with solid lines.}
\label{col_dens}
\end{figure}

\clearpage
\begin{figure}
\label{CorrInTime}
\includegraphics[width=0.4\textwidth,clip=,angle=90]{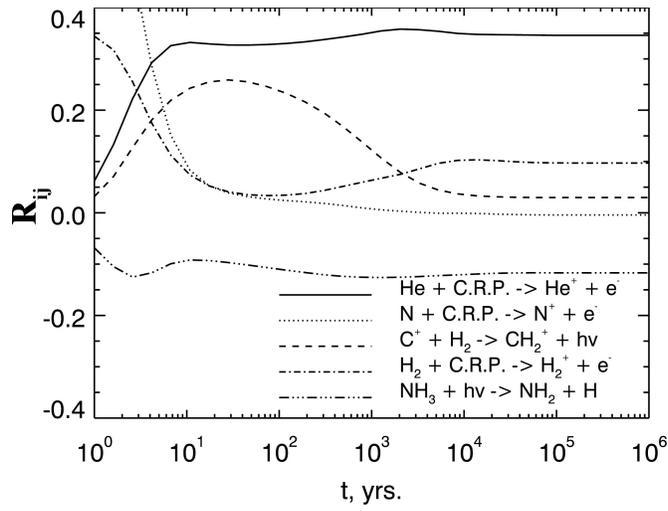}
\caption{The largest correlation coefficients of
relevant reactions for the evolution of NH$_3$ as a function of time
in the disk intermediate layer ($r=97$~AU, $z=30$~AU).}
\end{figure}


\clearpage
\begin{deluxetable}{ll}
\tablewidth{0pt} 
\tablecaption{Initial abundances\label{init_abund}}
\tablehead{\colhead{Species} & \colhead{n(X)/n(H)}}
\startdata
He    & 9.75(-2)$^{\rm a}$ \\
H$_2$ & 4.99(-1) \\
H   & 2.00(-3) \\
C   & 7.86(-5) \\
N   & 2.47(-5) \\
O   & 1.80(-4) \\
S   & 9.14(-8) \\
Si  & 9.74(-9) \\
Na  & 2.25(-9) \\
Mg  & 1.09(-8) \\
Fe  & 2.74(-9) \\
P   & 2.16(-10)\\
Cl  & 1.00(-9) \\
F  & 1.00(-10) \\
\enddata
\tablenotemark{a}\tablenotetext{a}{A(-B) means A$\times $10$^{-B}$}
\end{deluxetable}

\clearpage
\begin{deluxetable}{lll}
\tablewidth{0pt} 
\tablecaption{Typical column density uncertainties \label{abs_uncer}}
\tablehead{\colhead{Species} & \colhead{Standard} & \colhead{Improved} \\
\colhead{} & \colhead{model} & \colhead{model}
}
\startdata
H$_3^+$  & factor $\sim$2   &  $\sim$1.5 \\
C$^+$    & factor $\sim$2      & $\sim$1.5 \\
CO       & factor $\sim$1.25 & 1.1 \\
CO$_2$   & factor $\sim$3.5   & $\sim$2.5 \\
CS       & factor 3     & 2.5 \\
HCO$^+$  & factor $\sim 3$  & $\sim$1.5 \\
H$_2$CO  & factor 3   & $\sim$1.5 \\
CN       & factor 4 & 2 \\
HCN      & factor $\sim 3$  & 2.5 \\
HNC      & factor 3   & $\sim2.5$ \\
N$_2$H$^+$ & factor 2.5 & 2 \\
NH$_3$   & factor 2.5    & 2.5 \\
HCNH$^+$ & factor 2.5      & 2.5 \\
H$_2$O   & factor 1.7    & 1.4 \\
C$_2$H   & factor 4       & $\sim 3$\\
\enddata
\end{deluxetable}

\clearpage
\begin{deluxetable}{llllll}
\tablewidth{0pt}
\tabletypesize{\footnotesize} 
\tablecaption{Most problematic reactions\label{imp_reac}}
\tablehead{\colhead{Reaction} & \colhead{} &\colhead{} & \colhead{Uncertainty} & \colhead{Weight} & \colhead{Type}  }
\startdata
{\bf H$_2$ +  C } & {\bf $\rightarrow$ }       & {\bf CH$_2$ +  h$\nu$ }    & {\bf factor 10 } &  1.0  & {\bf RA } \\
{\bf H$_2$ + C$^+$} & {\bf $\rightarrow$ }  & {\bf CH$_2^+$ + h$\nu$ } & {\bf factor 10 } &  4.2(-1)  & {\bf RA } \\
H$_2$  +     CH     & $\rightarrow$   &   CH$_3$  +      $h\nu$           & factor 10 & 1.6(-2)  & RA \\
H$_2$  +      CH$_3^+$   & $\rightarrow$   &   CH$_5^+$  +     $h\nu$           & factor 10 & 7.6(-2)  & RA \\
C   +       H      & $\rightarrow$   &   CH   +      $h\nu$           & factor 10 & 2.3(-2)  & RA \\
C   +       N      & $\rightarrow$   &   CN   +      $h\nu$           & factor 10 & 2.2(-2)  & RA \\
C$^+$   +       e$^-$  & $\rightarrow$   &   C    +      $h\nu$           & factor 10 & 1.2(-2)  & RR \\
CH$_3^+$ +  e$^-$  & $\rightarrow$   &   CH$_3$ +     $h\nu$           & factor 10 &  1.2(-2) &RR \\
& & & \\
{\bf H$_2$ +  C.R.P.} & {\bf $\rightarrow$ } & {\bf H$_2^+$  + e$^-$ } & {\bf factor $2^{\rm *}$ } & 1.4(-1)& {\bf CRI }  \\
H$_2$ +     C.R.P.    & $\rightarrow$   &   H$^+$   +      H     +     e$^-$    & factor 2 & 1.0(-2)  & CRI \\
He    +     C.R.P.    & $\rightarrow$   &   He$^+$   +     e$^-$               & factor 2 & 2.3(-1)  & CRI \\
N     +     C.R.P.    & $\rightarrow$   &   N$^+$   +      e$^-$               & factor 2 & 9.7(-2)  & CRI \\
& & & \\
C   +       $h\nu$ & $\rightarrow$   &   C$^+$  +       e$^-$               & factor 2 & 3.8(-1)  & PI \\
C   +       C.R.P.$h\nu$ & $\rightarrow$   &   C$^+$   +      e$^-$         & factor 2 & 6.7(-2)  & PI \\
CO  +       $h\nu$ & $\rightarrow$   &   O    +      C                & factor 10 & 5.6(-2)  & PD \\
CH  +       $h\nu$ & $\rightarrow$   &   C    +      H                & factor 2 & 2.3(-2)  & PD \\
CH$_2$  +   $h\nu$ & $\rightarrow$   &   CH   +      H                & factor 2 & 6.8(-3)  & PD \\
CN  +       $h\nu$ & $\rightarrow$   &   N    +      C                & factor 2 & 3.2(-2)  & PD \\
CS   +      $h\nu$ & $\rightarrow$   &   S   +       C                & factor 2 & 2.6(-2)  & PD \\
NH$_3$ +    $h\nu$ & $\rightarrow$   &   NH$_2$  +   H                & factor 1.5 & 8.2(-3)  & PD \\
HCN   +     $h\nu$ & $\rightarrow$   &   CN   +      H                & factor 1.5 & 1.5(-2)  & PD \\
OH  +       $h\nu$ & $\rightarrow$   &   O    +      H                & factor 1.5 & 7.8(-3)  & PD \\
& & & \\
He$^+$    +     CO      & $\rightarrow$   &   O + C$^+$   +      He                & factor 1.25 & 9.6(-3)  & IN \\
He$^+$    +     N$_2$   & $\rightarrow$   &   N + N$^+$   +      He                & factor 1.25 & 7.4(-3)  & IN \\
H$_3^+$  +      C      & $\rightarrow$   &   CH$^+$   +     H$_2$               & factor 2 & 3.3(-2)  & IN \\
H$_3^+$  +      O      & $\rightarrow$   &   OH$^+$   +     H$_2$               & factor 1.5 & 1.0(-2)  & IN \\
H$_3^+$  +      CO     & $\rightarrow$   &   HCO$^+$   +     H$_2$               & factor 1.25 & 7.3(-3)  & IN \\
H$_2^+$  +      N      & $\rightarrow$   &   NH$^+$   +     H                & factor 2 & 4.3(-2) & IN \\
H$_2$    +     He$^+$    & $\rightarrow$   &   He   +      H$^+$    +     H      & factor 2 & 2.0(-2)  & IN \\
H$_2$   +      He$^+$    & $\rightarrow$   &   He   +      H$_2^+$              & factor 2 & 2.1(-2)  & IN \\
H$_2$   +      NH$_3^+$   & $\rightarrow$   &   NH$_4^+$   +    H                & factor 1.5 & 4.4(-2)  & IN \\
H$^+$    +     O      & $\rightarrow$   &   O$^+$   +      H                & factor 1.5 & 8.4(-3)  & CT \\
C$^+$   +      CH     & $\rightarrow$   &   C$_2^+$   +     H                & factor 2 & 1.3(-2)  & IN \\
C$^+$   +      OH     & $\rightarrow$   &   CO$^+$    +    H                & factor 2 & 3.1(-2)  & IN \\
C$^+$   +      NH     & $\rightarrow$   &   CN$^+$   +     H                & factor 2 & 1.1(-2)  & IN \\
C$^+$   +      HCN    & $\rightarrow$   &   CNC$^+$   +     H                & factor 1.25 & 7.9(-3)  & IN \\
C     +     HCO$^+$   & $\rightarrow$   &   CO    +     CH$^+$              & factor 2 & 2.7(-2)  & IN \\
CH    +     S$^+$     & $\rightarrow$   &   CS$^+$   +     H                & factor 2 & 2.6(-2)  & IN \\
CH$_2^+$    +     O     & $\rightarrow$   &   HCO$^+$   +     H                & factor 2 & 7.2(-3)  & IN \\
NH$_3^+$  +   Mg      & $\rightarrow$   &   NH$_3$   +     Mg$^+$           & factor 2 & 1.1(-2)  & CT \\
& & & \\
C    +      CH$_2$    & $\rightarrow$   &   C$_2$H   +     H                & factor 2 & 3.6(-2)  & NN \\
C    +      C$_2$H    & $\rightarrow$   &   C$_3$    +     H                & factor 2 & 8.3(-3)  & NN \\
N    +      CH$_2$    & $\rightarrow$   &   HCN  +      H                & factor 1.5 & 8.5(-2)  & NN \\
N    +      CN     & $\rightarrow$   &   N$_2$   +      C                & factor 2 & 4.6(-2)  & NN \\
N    +      HCO    & $\rightarrow$   &   HCN   +     O                & factor 2 & 2.9(-2) & NN \\
H    +      CH$_2$    & $\rightarrow$   &   CH   +      H$_2$               & factor 1.25 & 1.9(-2)  & NN \\
H     +     CH     & $\rightarrow$   &   C    +      H$_2$               & factor 1.5 & 1.1(-2)  & NN \\
CH   +      O      & $\rightarrow$   &   HCO$^+$  +     e$^-$               & factor 1.5 & 4.0(-2)  & NN \\
CH   +      O      & $\rightarrow$   &   CO  +     H                 & factor 1.25 & 8.3(-3) & NN \\
CH   +      N      & $\rightarrow$   &   CN    +     H                & factor 1.25 & 2.3(-2)  & NN \\
CH    +     S      & $\rightarrow$   &   CS    +     H                & factor 2 & 1.6(-2)  & NN \\
CH$_2$  +      S      & $\rightarrow$   &   CS    +     H$_2$               & factor 2 & 6.3(-2)  & NN \\
CH$_2$  +      O      & $\rightarrow$   &   CO    +     H   +       H      & factor 1.25 & 1.0(-2)  & NN \\
CH$_3$  +      O      & $\rightarrow$   &   H$_2$CO   + H                  & factor 1.25 & 1.4(-2)  & NN \\
& & & \\
HCO$^+$   +    e$^-$     & $\rightarrow$   &   CO   +      H                & factor 1.25 & 3.1(-2)  & DR \\
NH$_4^+$   +    e$^-$     & $\rightarrow$   &   NH$_3$ + H                   & factor 1.25 &  8.1(-3)  & DR \\
\enddata
\tablenotemark{*}\tablenotetext{*}{These rates are partly uncertain
due to uncertainties in physical parameters.} \tablecomments{(CRI)
Cosmic ray ionization; (CT) Charge transfer; (DR) Dissociative
recombination; (IN) ion-neutral reaction; (NN)
Neutral-neutral reaction; (PD) Photodissociation; (PI)
Photoionization; (RA) Radiative association; (RR) Radiative
recombination. Three most important reactions for the disk chemistry
as identified by the sensitivity analysis are written in boldface.}
\end{deluxetable}

\end{document}